\documentclass{article}

\usepackage{amsmath}
\usepackage{amssymb}

\usepackage{graphicx}

\usepackage{cite}
\usepackage{color} 

\usepackage[labelfont=bf,labelsep=period,justification=raggedright]{caption}

\makeatletter
\renewcommand{\@biblabel}[1]{\quad#1.}
\makeatother

\date{}

 \newcommand{\apgt}{\ {\raise-.5ex\hbox{$\buildrel>\over\sim$}}\ }
 \newcommand{\aplt}{\ {\raise-.5ex\hbox{$\buildrel<\over\sim$}}\ }

\begin{document}

% Title must be 150 characters or less
\begin{flushleft}
{\Large
\textbf{Evidence of strategic periodicities in collective conflict dynamics}
}
% Insert Author names, affiliations and corresponding author email.
\\
Simon DeDeo$^{1,\ast}$, 
David Krakauer$^{1}$, 
Jessica Flack$^{1,2}$ 
\\
\bf{1} Santa Fe Institute, 1399 Hyde Park Road, Santa Fe, NM 87501
\\
\bf{2} Yerkes National Primate Research Center, Emory University, Atlanta, GA 30322

$\ast$ E-mail: simon@santafe.edu
\end{flushleft}

% Please keep the abstract between 250 and 300 words
\section{Abstract}
We analyze the timescales of conflict decision-making in a primate society. We present evidence for multiple, periodic timescales associated with social decision-making and behavioral patterns. We demonstrate the existence of periodicities that are not directly coupled to environmental cycles or known ultraridian mechanisms. Among specific biological and socially-defined demographic classes, periodicities span timescales between hours and days, and many are not driven by exogenous or internal regularities. Our results indicate that they are instead driven by strategic responses to social interaction patterns. Analyses also reveal that a class of individuals, playing a critical functional role, policing, have a signature timescale on the order of one hour. We propose a classification of behavioral timescales analogous to those of the nervous system, with high-frequency, or $\alpha$-scale, behavior occurring on hour-long scales, through to multi-hour, or $\beta$-scale, behavior, and, finally $\gamma$ periodicities observed on a timescale of days. 

\section{Introduction}
 \label{methods}

Variability on multiple timescales is a fundamental feature of complex systems \cite{Sompolinsky:1981p18132,Coolen:1994p18126}. Minimally, multiple timescales are critical for feedback, and without them there would be no memory, regulation, or adaptation. Adaptation, for example, requires timescales fast relative to the environment. Memory, on the other hand, arises from slow variables that average over the underlying fast dynamics. These slow variables can serve as a reference for decision-making when the lower-level dynamics are rapidly fluctuating \cite{Flack2007,Kiebel2009, Boehm2010,Flack2010}.

In the brain, multiple timescales, or characteristic frequencies of oscillation \cite{Buzsaki:2004p17637}, enable populations of neurons to efficiently represent different kinds of statistical information about the environment. Timescales have been hypothesized to play a role in the emergence of a unitary consciousness by binding the activity of large populations of cells \cite{gazzaniga2004} and to provide, by increasing the combinatorial space, new means of storing complex temporal patterns \cite{Kiebel2009} and optimizing synaptic weights.

Timescale variability has also been observed in behavioral dynamics and social systems. The dynamics of learning (\emph{e.g.}, \cite{Newell2001}) and decision-making (\emph{e.g.}, \cite{selezneva2006}) occur at timescales from seconds through to months and years. Social systems are comprised of emergent, hierarchically organized social networks that change over a broad range different timescales from hours to years \cite{Boehm2010,Flack2010}. These observations raise interesting questions including, how new timescales emerge, and what the optimal coupling is between time constants given functional requirements at the individual and collective levels. To answer these questions we must first quantitatively characterize the range of time constants in our study systems. Whereas much is known about the range of time constants in neural systems, there is has been less quantitative characterization of time constants and their implications in social phenomena.

Here we show that conflict decision-making behavior -- specifically, the decision to join fights -- in a primate society is characterized by multiple, periodic timescales. We report the range of timescales detected, and propose a broad classification of behavioral time scales into $\alpha$ waves on hour scales, $\beta$ waves on multi-hour scales, and $\gamma$ waves from six hours up to days. We find that the timescales we detect are properties of demographic classes defined by biological properties, like age and sex, or social properties, like power and social roles.  

Our analyses take as input a well-studied conflict timeseries\cite{DeDeo:2010p18133} collected from a large, socially-housed, primate group (pigtailed macaques, \emph{Macaca nemestrina}) at the Yerkes National Primate Center in Lawrenceville, Georgia (see Sec. \ref{Data}). To characterize the range of timescales in our study system, we adapt a technique developed to study irregularly-sampled astronomical phenomena, the Lomb-Scargle periodogram.  As described in Sec. \ref{LSM}, the Lomb-Scargle periodogram can be used to detect a very important class of timescales generated by regular periodicities in the dynamics. Examples of phenomena with these kinds of periodicities include the ultraridian waves of physiology, circadian rhythms correlated with the photoperiod, and seasonal and reproductive infraridian periodicities \cite{Foster}. 

Using the Lomb-Scargle periodogram, we extract signatures of broad-band variation from the conflict time-series. We consider three alternative hypotheses to account for the time constants we observe. These include two null models intended to determine whether the timescales are the consequence of exogenous or endogenous drivers of behavior, and a third hypothesis: the timescales are generated by strategically timed decisions to join or avoid fights. By ``strategic'' we mean the decision to join or avoid fights is timed in response to the pattern of social interactions rather to external cues or physiological clocks (see Sec. \ref{diurnal} for an operational definition on what is meant by ``strategy''). 

\section{Description of Lomb-Scargle Method}
\label{LSM}

The quantitative study of timescale variation falls under the heading of \emph{spectral analysis}. A tool ideally suited to the spectral analysis of sparsely and irregularly sampled data is the Lomb-Scargle periodogram \cite{Lomb:1976p17776,Scargle:1982p17636,press1996}. One of the advantages of irregularly sampled data is the great reduction in windowing and aliasing effects \cite{press1996}.

The Lomb Periodogram for a time-series, $\{h_{j}\}$, $j=1\ldots N$, sampled at times $\{t_{j}\}$, is defined by 
\begin{equation} 
P(\omega) \equiv \frac{1}{2Z}\left(\frac{\left[\sum_{j}(h_{j}-\bar{h})\cos{\omega(t_{j}-\tau)}\right]^{2}}{\sum_{j}\cos^{2}{\omega(t_{j}-\tau)}}+\frac{\left[\sum_{j}(h_{j}-\bar{h})\sin{\omega(t_{j}-\tau)}\right]^{2}}{\sum_{j}\sin^{2}{\omega(t_{j}-\tau)}}\right),
\end{equation}  
where $Z$ is a normalization and $\tau$ is defined by 
\begin{equation} \tan{2\omega\tau}=\frac{\sum_{j}\sin{2\omega t_{j}}}{\sum_{j}\cos{2\omega t_{j}}}.\end{equation}  
As discussed in Ref. \cite{press1996}, if we write \begin{equation} h(t)=A\sin{\omega t}+B\cos{\omega t},\end{equation}
the definition of $\tau$ amounts to setting $P(\omega)$ proportional to $A^{2}+B^{2}$, where the coefficients are set by a linear least-squares fit. Choosing the normalization, $Z$, to be the variance of $\{h_{j}\}$ makes the estimator the Lomb-Scargle normalized periodogram. A nice feature of the periodogram is that, under a null hypothesis of i.i.d. Gaussian variables, the distribution of $P(\omega)$ is an exponential distribution with mean unity \cite{Scargle:1982p17636}.

The output of the Lomb-Scargle is a (normalized) strength-of-signal as a function of frequency, $P(\omega)$. The frequency, $\omega$, measured in Hz, is simply the inverse period (times $2\pi$). In our case, the different $\{h_j\}$ for which we measure $P(\omega)$ are associated with conflicts involving different individuals and demographic classes, as described in the following section.

\section{Structure of the Data}
\label{Data}
Our study group contained 48 socially-mature individuals and 84 individuals in total. Conflicts, or `fights,' in this system involve two or more individuals and are separated by peaceful periods -- defined as the absence of fights among any of the group members. Operational definitions, and additional details on the data set and data collection protocol appear in the Materials and Methods. 

Briefly, a ``fight'' was operationally defined to include any interaction in which one individual threatens or aggresses a second individual. A conflict was considered terminated if no aggression or withdrawal responses (fleeing, crouching, screaming, running away, submission signals) was exhibited by any of the conflict participants for \emph{two minutes} from the last such event. Fights involve multiple individuals, ranging in size from two to twenty-eight individuals. Fights can be conceived of as small networks that grow and shrink as pair-wise and triadic interactions become active or terminate, until there are no more individuals fighting under the above-described two minute criterion. As described in the Methods (Sec. \ref{op}, only data on time of fight onset and the individuals involved in the fight are used in these analyses. No time data are available within fights; although the order of an individualÕs entry was noted during data collection, this information was not used in our analyses. Fight onset and termination time (using above-described criterion) were noted in hours, minutes, and seconds (see Sec. \ref{op} for further detail).

Our interest is in whether timing influences the decision of an individual to join fights. However, because the average number of fights per individual is low, it is hard to detect a signal using the Lomb-Scargle periodogram at the individual level. Hence for most of our analyses, we aggregate individuals into demographic classes, according to biological and social characteristics, and ask whether, taken collectively, individuals of a given class exhibit a timescale on which they join fights. 

The biological classes we consider include age-class (socially-mature individuals, and two subclasses of the socially-mature set: subadults and adults), sex, and matriline (female and all daughters one year of age or older). The criteria we use to define social classes -- social power \cite{Flack2006b} and performance of policing role -- have been shown in previous work to be important factors in structuring social interactions in the study group \cite{Flack2005, Flack2006}). Demographic class sample sizes are provided with each analysis. For further details on these demographic classes and for definitions of power and policing, see Materials and Methods. 

We calculate $P(\omega)$ for each of the demographic classes described above. For each demographic class, the Lomb-Scargle Periodogram takes as input a discrete series of measurements from the conflict time series. The timing of an event, $t_i$, is set to be the onset of a fight in the observations; the $h_i$ is a discrete variable: the presence (1) or absence (0) of an individual, or, in the case of classes, the number of individuals involved in a fight at $t_i$ from that demographic class. In effect, a conflict is considered to ``sample'' the dispositions of  demographic class in question. \emph{Detection of a signal using the Lomb-Scargle periodogram indicates a timescale on which members of that class join or avoid fights.}

Conflicts are short, with a median duration of only 15 seconds. The scales we recover span nearly six orders of magnitude in timescales -- between tens and and tens of millions of seconds. Of that range, the range of scales between $10^{3}$ seconds (tens of minutes) and $10^{5}$ seconds (days) is most accessible. On much longer scales, measurements of nearby periodicities are strongly correlated, meaning there are few independent measurements to be made. Meanwhile, on the very shortest scales, the finite duration of conflicts tends to wash out signals. 

In previous work \cite{DeDeo:2010p18133}, we found evidence that the decision to join fights made by individuals in this study group depended on the properties of the preceding conflict event. The median time between conflicts is 255 seconds and so decisions to join conflicts correspond to the shortest timescales accessible to our analysis. Ref. \cite{DeDeo:2010p18133} tested a set of alternative causal strategies, or behavioral production rules -- formally denoted as $\mathcal{C}(n,m)+$\texttt{AND/OR} -- that could be giving rise to these time scales, and found $\mathcal{C}(2,1)+$\texttt{AND} to be a dominant strategy. This indicates that an individual decides to join the the current conflict because a specific pair of individuals appeared in the previous conflict. This rule applies to all individuals the group. As many, though not all, adjacent fights are separated by only a few minutes, this finding suggests that second-to-minute reasoning scales can be of great importance to conflict dynamics. 

Can timescales longer than this be found directly in this ruleset? Because any particular rule is invoked so infrequently, detecting shifts becomes difficult, if not impossible. The analysis we present in this paper does not need to measure whether individual decision-making rules change, and so does not suffer from the same signal-to-noise issues.

\section{Results}
As described in Sec.~\ref{LSM}, the output of the Lomb-Scargle method is a plot -- a periodogram -- of the fluctuation power as a function of frequency (or inverse period). Power at a particular frequency or range of frequencies indicates the presence of fluctuations with those characteristic timescales. For example, if an individual or demographic class is characterized by a tendency to shift behaviors (from, for example, less conflict prone to more conflict prone and back) on timescales of roughly one hour, one would see an above the null model bump in the periodogram in that range (see Fig.~\ref{adults}, for an example of a periodogram with significant detections in a number of logarithmically-spaced bins.)

As in previous work, the highly correlated nature of the system means the choice of adequate null models is crucial. In presenting our results, we consider two null models: the mixed-strategy null, and a stronger, daily-forcing null. The former, discussed in Sec.~\ref{mixed}, looks for signatures of changing behavioral dispositions; the latter, discussed in Sec.~\ref{diurnal}, tries to explain these changes by a model of context-insensitive daily shifts in conflict behavior. Features unexplained by either null -- and associated with decision-making that is sensitive to fluctuations about mean behavior -- are of particular interest. We focus on them in Sec.~\ref{strategic}. 

In analyses such as these, where many bins are searched for signal, a distinction arises between the statistical significance of a single-bin detection, and the statistical significance of the detection overall. As an extreme example, if one searches one hundred bins, each considered equally likely to harbor a signal, a $p$ value of $10^{-2}$ in a single bin does not imply an overall significant detection. In some cases, such as the subadult male demographic class (see Fig. ~\ref{sex}) a single bin shows a strong above-null detection, but combining that $p$ value with many non-detections in other bins reduces the significance.

\subsection{The Mixed-Strategy Null} \label{mixed}

Without strong priors on timescales of variability -- or an intuition for the expected signal strength in the periodogram -- we first compare the observed variability against a null model that retains only the time-independent properties of the data. 

We produce a set of null periodograms by shuffling the time series. We keep the timing of fights the same, but shuffle their internal compositions, so a fight at time $t$ in the data will correspond to a fight at time $t$ in all null sets, but will have the composition of a different fight, drawn (without replacement) from a different time $t^{\prime}$. The normalization of the Lomb-Scargle periodogram is such that the mean value of the null is unity; further statistical issues associated with null model estimation are discussed in the Materials and Methods.

In a game theoretic context, this null model corresponds to assigning individuals a constant mixed strategy (in the game-theoretic sense of mixed): time-independent probabilistic play of one of two strategies, ``join conflict'' or ``avoid.'' Note that  Lomb-Scargle analyses the data in terms of periodic functions; failure to reject the null suggests that the animals are playing probabilistic strategies without strong periodicities. 

In the case of a demographic class of size $n$, the equivalent mixed strategy is for the group as a whole, and amounts to a probabilistic choice of $n+1$ options -- ``none of us join,'' ``one of us joins,'' and so forth to ``all $n$ of us join.'' This is a distinct process from averaging, over a demographic class, the periodograms obtained for individuals. It is sensitive to the timescales for \emph{collective} behaviors of a demographic class, which may be different from the timescales of its individuals. For example, in the case that two individuals in a demographic class alternate their participation -- perhaps because either is sufficient to play a particular functional role -- the observed timescale for the group will be faster, and more coherent, than that of either of the two individuals taken independently.

Evidence for non-null behavior indicates failure of the assumption of stationary and memoryless play; it is, among other things, \emph{prima facie} evidence against the convergence to a stationary solution concept  \cite{fudenberg1998} -- unless, of course, the ``game'' is assumed to take place on timescales longer than those detected in the data. 

In behavioral terms, this choice of null allows us assess whether there are non-stationary features of behavior over and above static properties that tell us about an individual's or demographic class' overall willingness to engage in conflict. 

Additionally, as discussed in the Materials and Methods, the null allows us to bound the influence of systematic effects, due to the sampling strategy or the correlations induced by noise, that might affect naive estimates of statistical significance. 

\begin{figure} \includegraphics[width=3.275in]{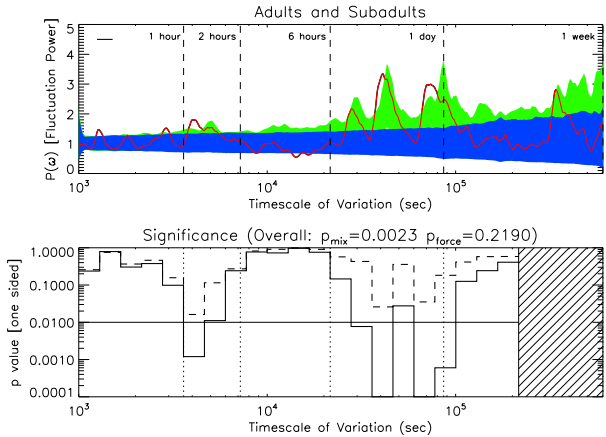} \caption{Timescales of the decision to join fights for the socially-mature individuals considered as a demographic class ($n=47$). Top panel: Lomb-Scargle Periodogram for the socially-mature demographic class. The data are shown as the solid red line. The (darker) blue band shows the $p=0.05$ confidence for the mixed-strategy null of Sec.~\ref{mixed}; the (lighter) green band shows the $p=0.05$ confidence for the daily forcing null of Sec.~\ref{diurnal}. Bottom panel: one-sided $p$-value significance levels for the mixed (solid line) and daily (dashed line) null models, showing evidence for $\alpha$ and $\gamma$ oscillations, between $10^3$ seconds and $2.5$ days. (See Materials and Methods for further details). The overall significance of deviation from the mixed-strategy null is $p\sim10^{-3}$; the fluctuations are consistent with daily forcing.} \label{adults}\end{figure}

Of the 47 socially mature adults we consider in these analyses (see Methods), 6 show significant ($p<0.01$) deviations from the mixed-strategy null when their individual patterns of behavior are examined. By analyzing at the demographic class level, we increase our signal-to-noise and are able to detect significant patterns in the timescale spectra. 

Fig.~\ref{adults} shows the periodogram for the aggregated data of the 47 socially-mature individuals; the top panel shows the (smoothed) power at each timescale, whereas the bottom panel shows the significance of any above-null power. The $p$-values are computed for the two conceptually distinct null models. Strong signals at two well-separated scales are the first evidence for timescales of behavior. The faster, $\alpha$, scale, is at one hour; whereas there are broad $\gamma$-scale oscillations  between eleven and twenty-four hours.

\begin{figure}
\begin{tabular}{cc}
\includegraphics[width=3.275in]{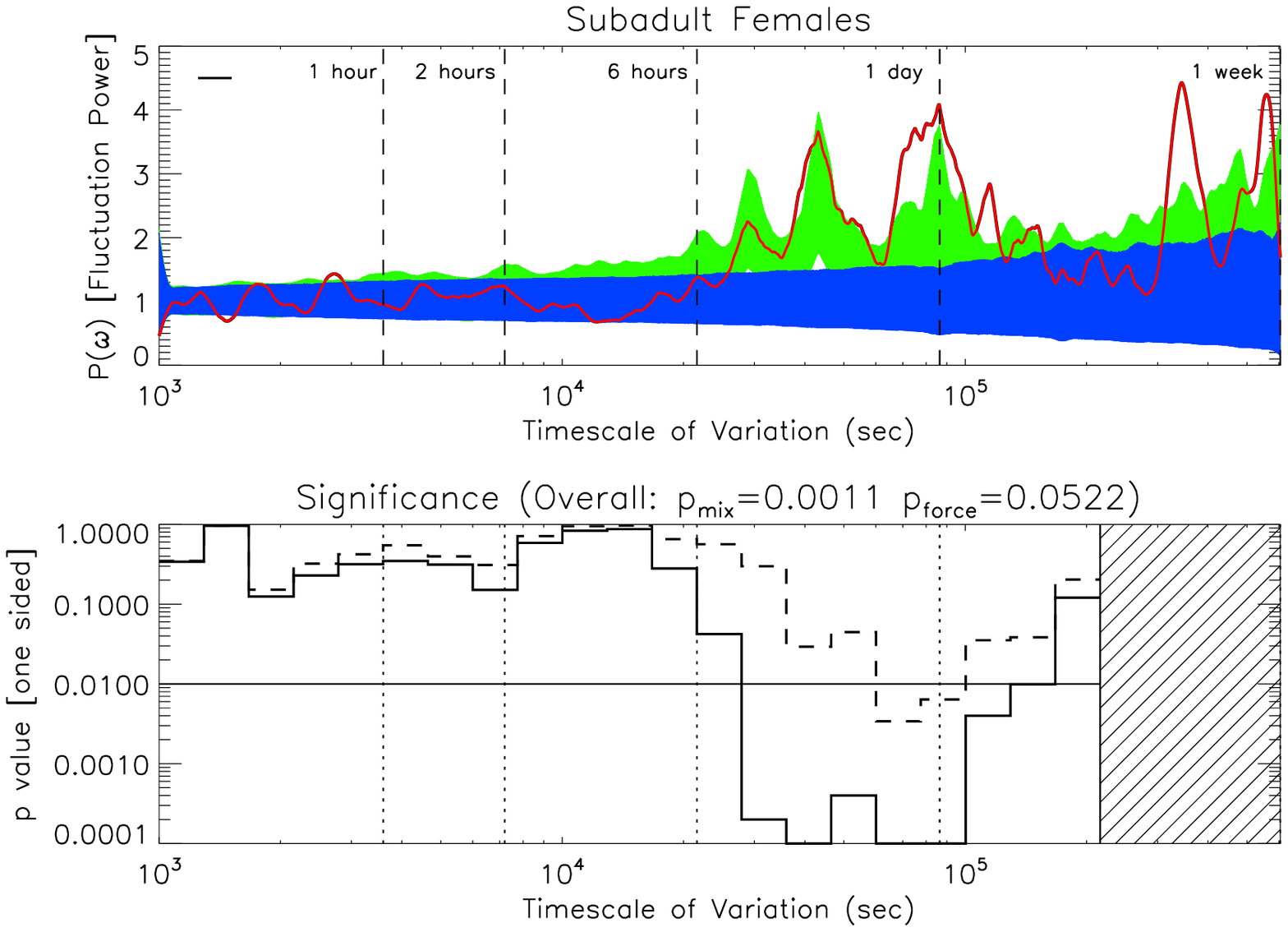} &
\includegraphics[width=3.275in]{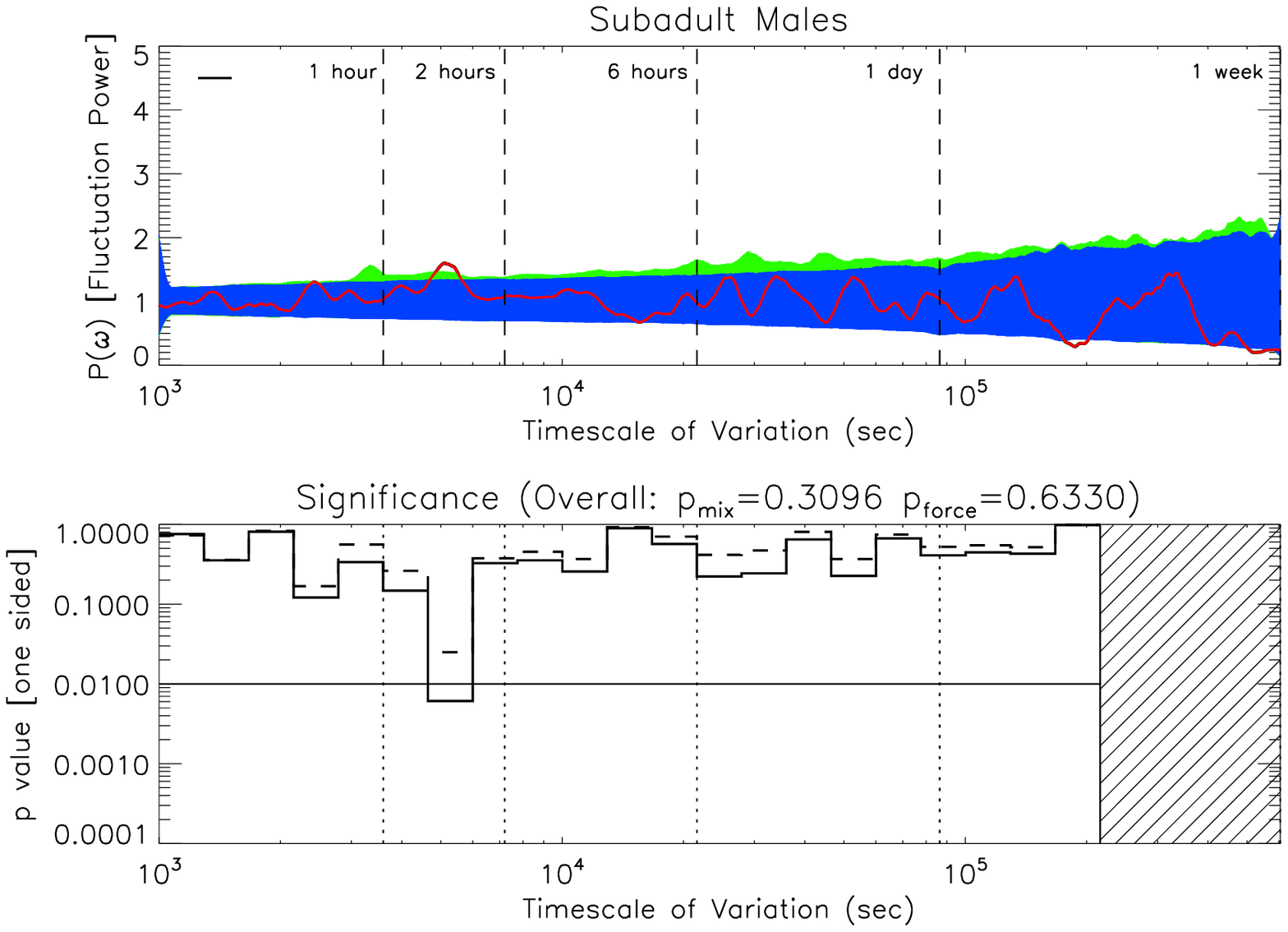} \\
\includegraphics[width=3.275in]{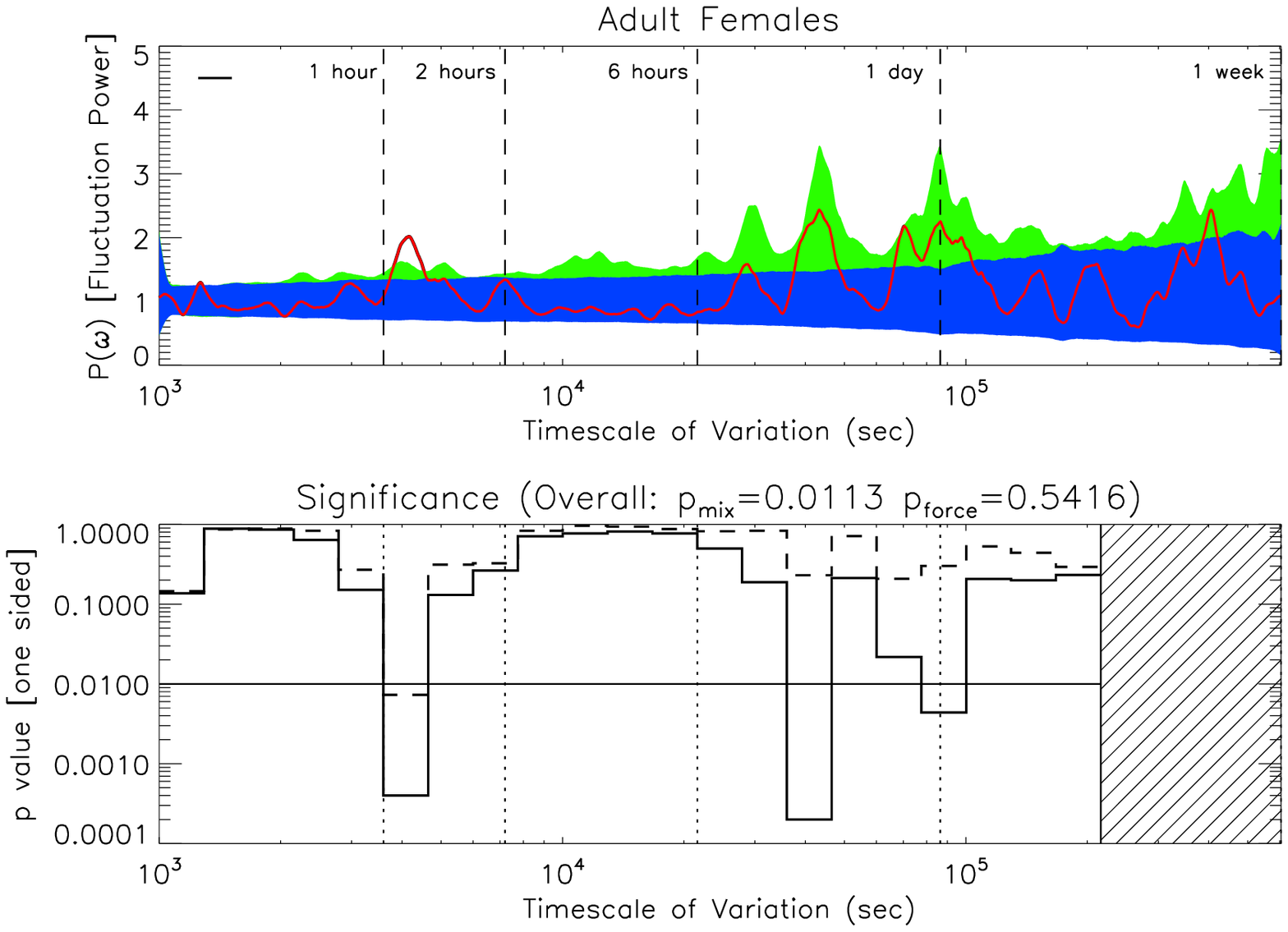} &
\includegraphics[width=3.275in]{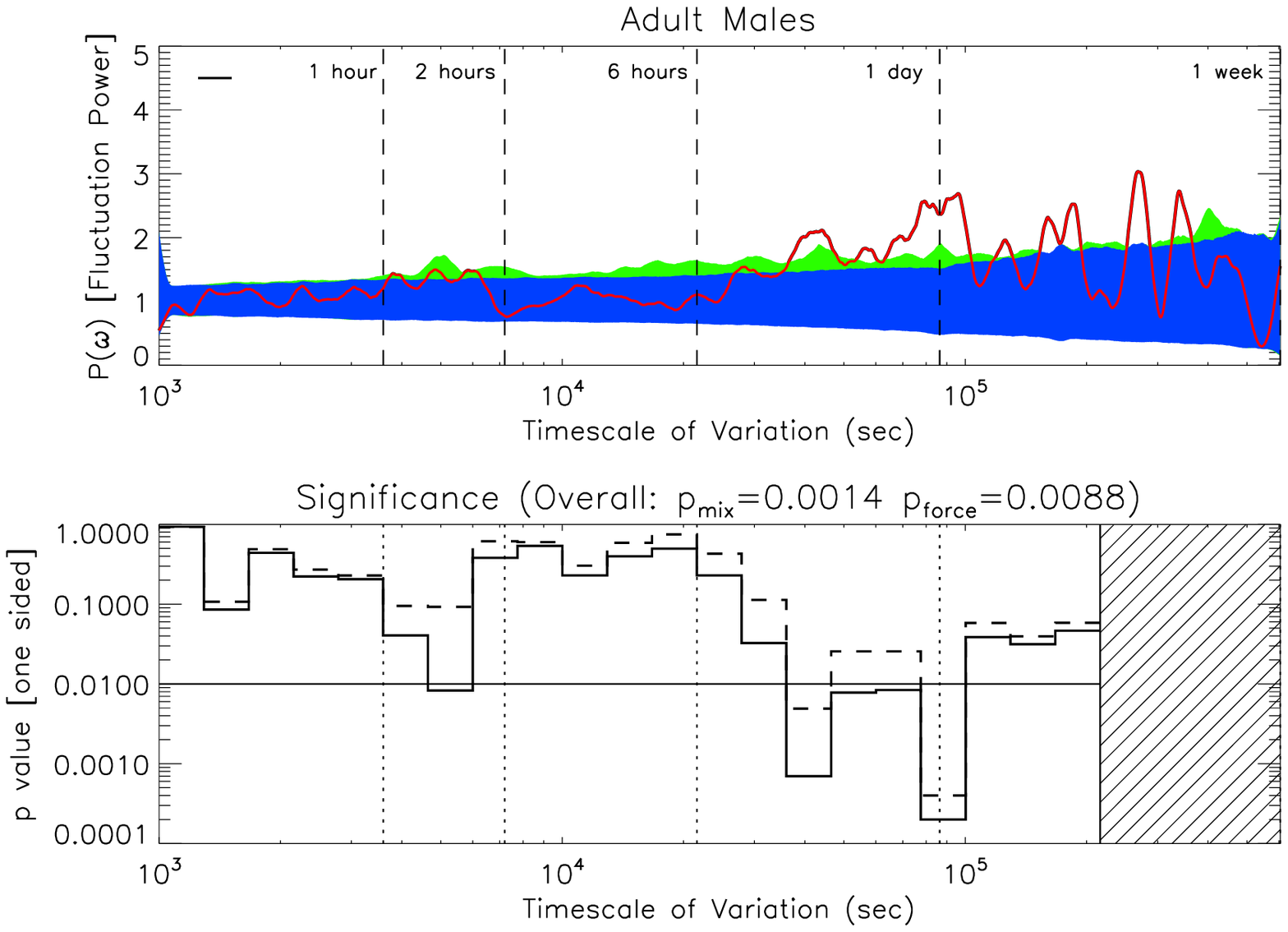}
\end{tabular}
\caption{Demographic classes defined by sex and age show different timescales. Top left pair: periodogram, and $p$-values, for the subadult females ($n=11$.) Bottom left pair: adult females ($n=22$.) Top right pair: the subadult males ($n=6$.) Bottom right pair: the adult males ($n=8$.) As before, the periodogram data are shown in red, and the blue band shows the $p=0.05$ confidence for the mixed-strategy null of Sec.~\ref{mixed}; the green band shows the $p=0.05$ confidence for the daily forcing null of Sec.~\ref{diurnal}. The $p$-values are one-sided, and for the mixed (solid line) and daily forcing (dashed line) nulls discussed in the text.} \label{sex}\end{figure}

As an example of the mixed-strategy null, Fig.~\ref{sex} shows the periodograms for two biologically-defined demographic classes -- the subadult females (36 to less than 48 months old), and the subadult males (48 to less than 60 months old). The blue bands show the $p=0.05$ confidence levels for the mixed-strategy null. Although these two demographic classes appear in a similar numbers of conflicts, in similar frequencies, the periodogram uncovers striking differences in their timescales.

Whereas the subadult males show some evidence for an $\alpha$-scale oscillation ($p\sim10^{-3}$ in a single bin), their overall behavior is consistent with the mixed-strategy null. The subadult females show strong $\gamma$ oscillations on scales between eight and twenty-four hours, with a number of bins with $p\ll10^{-3}$.  The subadult females, in particular, have an overall $p$-value, against the mixed-strategy null, of $p\lesssim10^{-3}$. 

The adult females (48 months and older) and adult males (60 months and older) show similarly distinct timescales. The adult females show a strong $\alpha$-scale oscillation that overlaps with the subadult males; they also show evidence for $\gamma$ oscillations. The adult males show $\gamma$ oscillations, as well as  ($p\sim0.01$) evidence for the faster timescales seen in the subadult males and adult females. The adult females show $\alpha$ and $\gamma$ oscillations; their $\alpha$ waves are similar to the subadult males; their $\gamma$ waves are slightly weaker than the subadult females (but still detectable.)

In addition to the sex/age-defined demographic classes, the demographic classes defined in terms of social power show important differences. Grouping individuals by power \cite{Flack2006} reveals additional complexities in the timescale structure of conflict decision-making. We also find important structure in the functionally-defined policing class (four high power individuals that perform the majority of effective policing interventions \cite{Flack2005}.

In particular, whereas the policing class shows similar $\alpha$ and $\gamma$ oscillations to all 47 socially-mature individuals considered collectively, the $\alpha$ band signal is  absent in the remaining eight individuals that make up the top power quartile. The second quartile in power shows no evidence for either of these scales. Instead, this second tier shows evidence for an intermediate $\beta$-scale oscillation around three hours.

Interestingly, there is far less evidence for timescales inherent to particular matrilines. Of the eleven matrilines present in the study group, only two show evidence for strong intrinsic timescales; these are shown in the Materials and Methods, Figs.~\ref{m1_supplementary} through~\ref{m3_supplementary}. Since members of these matrilines are naturally included in other demographic classes that do have strong timescales, lack of evidence for matriline-level timescales suggests that the timescales on which individuals within any particular matriline decide to join or avoid conflicts differ enough that, when taken collectively, the signals are washed out. We present the full results in the Materials and Methods.

\subsection{Daily Forcing} \label{diurnal}

In the previous section, we found evidence for multiple behavioral timescales in our data. These timescales show evidence for systematic modulation of the behavior of different individuals and demographic classes. What is the nature of such modulation? 

As the name suggests, the daily-forcing null is intended to capture shifts in behavior due to external or systematic internal cues that act, over the course of the day, identically from day to day. Such forcing might be generated by daily shifts in ambient temperature, by a regular feeding time, or by internal processes such as fatigue that naturally accrue over the course of a day. Hence, the daily-forcing null is much more demanding -- \emph{i.e.}, conservative -- on the time series than the mixed-strategy null, as it allows for temporal inhomogeneity. 

Observationally, the daily-forcing null is equivalent to a time-varying mixed strategy in which the variation is constrained to be the same from day to day. The variation of the mixed-strategy is measured, for the demographic class in question, from the data itself (see Materials and Methods). In our analysis, we allow the shifts to occur on timescales as fast as (but no faster than) 15 minutes -- sufficient to model, if possible, even the fastest, $\alpha$ scales. The null does not, of course, specify what particular processes leads to these daily shifts. It can be a combination of external, internal, and social factors (the behaviors of others that shift due to their own external and internal factors).  

Deviations from the daily forcing null can be accounted for in two ways. On the one hand, since the null has no day-to-day variations, deviations might indicate forcing on longer timescales such as \oe strus, or that learning is causing accumulated shifts in behavior. These effects would be visible, in the periodograms themselves, as strong signals beyond $10^{5}$~seconds. There is some evidence, in the subadult female demographic class (Fig.~\ref{sex}), for these longer scales; there, fluctuations at and above 24 hours reject the daily forcing null at $p\lesssim10^{-3}$, and the overall significance has $p\sim0.01$.

The other explanation for deviations from daily forcing is a breakdown in the assumption of independently-distributed behavior. In this case, signatures of variability, over and above the daily forcing null, are associated with context-sensitive, or \emph{strategic}, decision-making. Mathematically, rejection of this null would mean that shifts in the average behavior are not driving the system on these scales. Instead it is the system's correlated responses to \emph{fluctuations} about that average. Purely random fluctuations about the average will generate spurious timescales, but these are accounted for by the sampling-with-replacement, and so violations the daily-forcing null indicate correlations in those responses.

\begin{figure} 
\begin{tabular}{cc}
\includegraphics[width=3.275in]{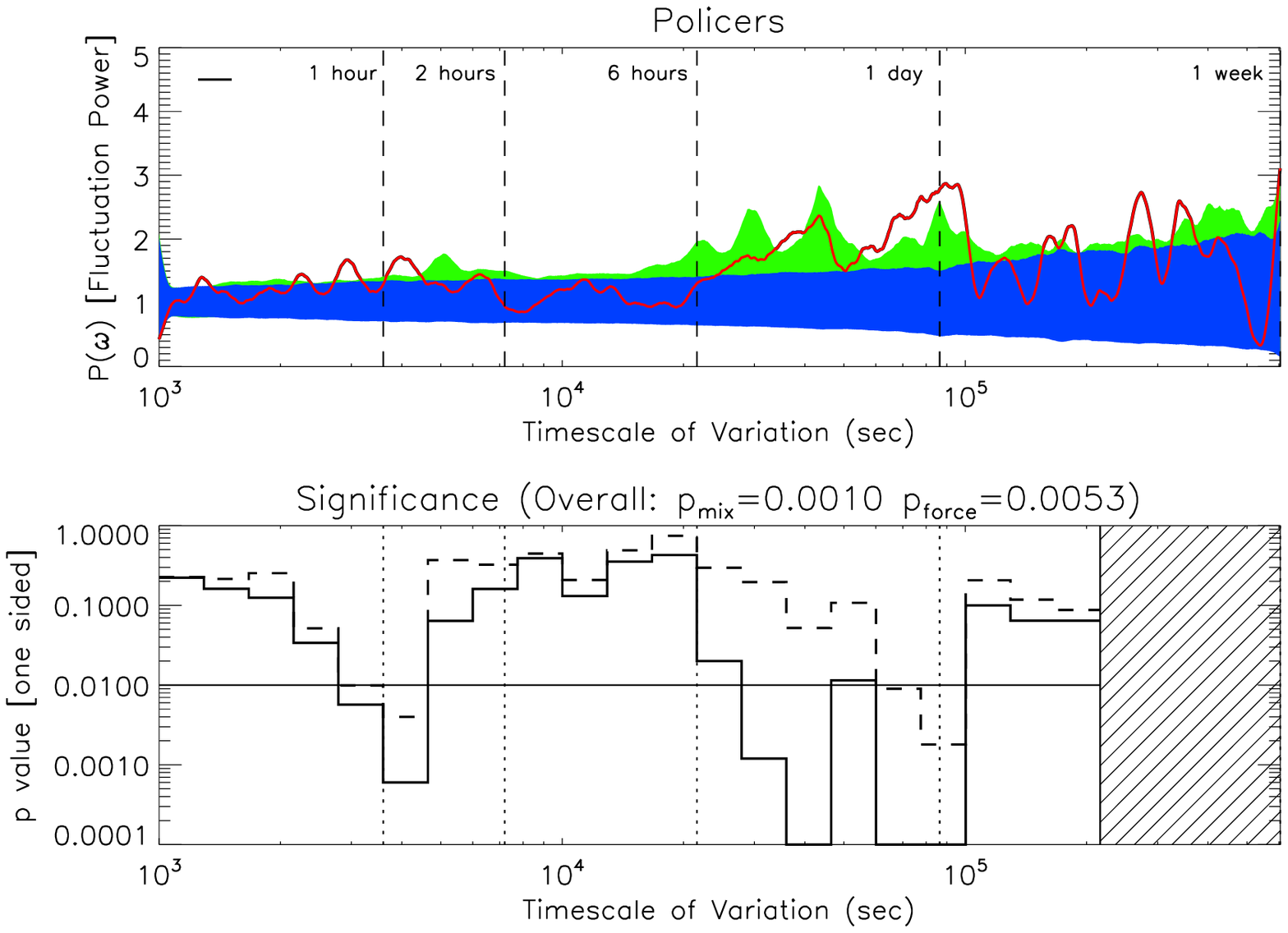} &
\includegraphics[width=3.275in]{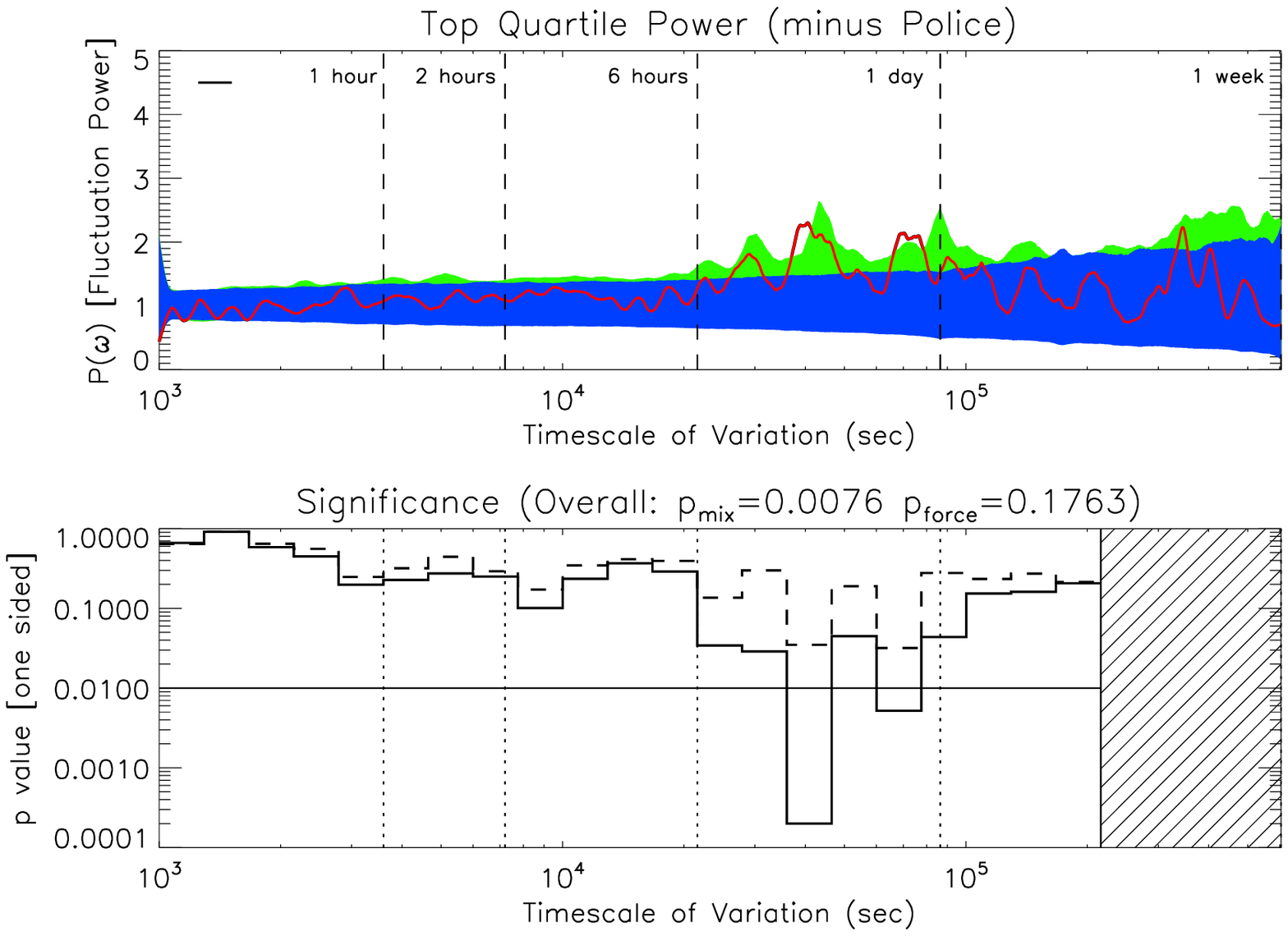} \\
\includegraphics[width=3.275in]{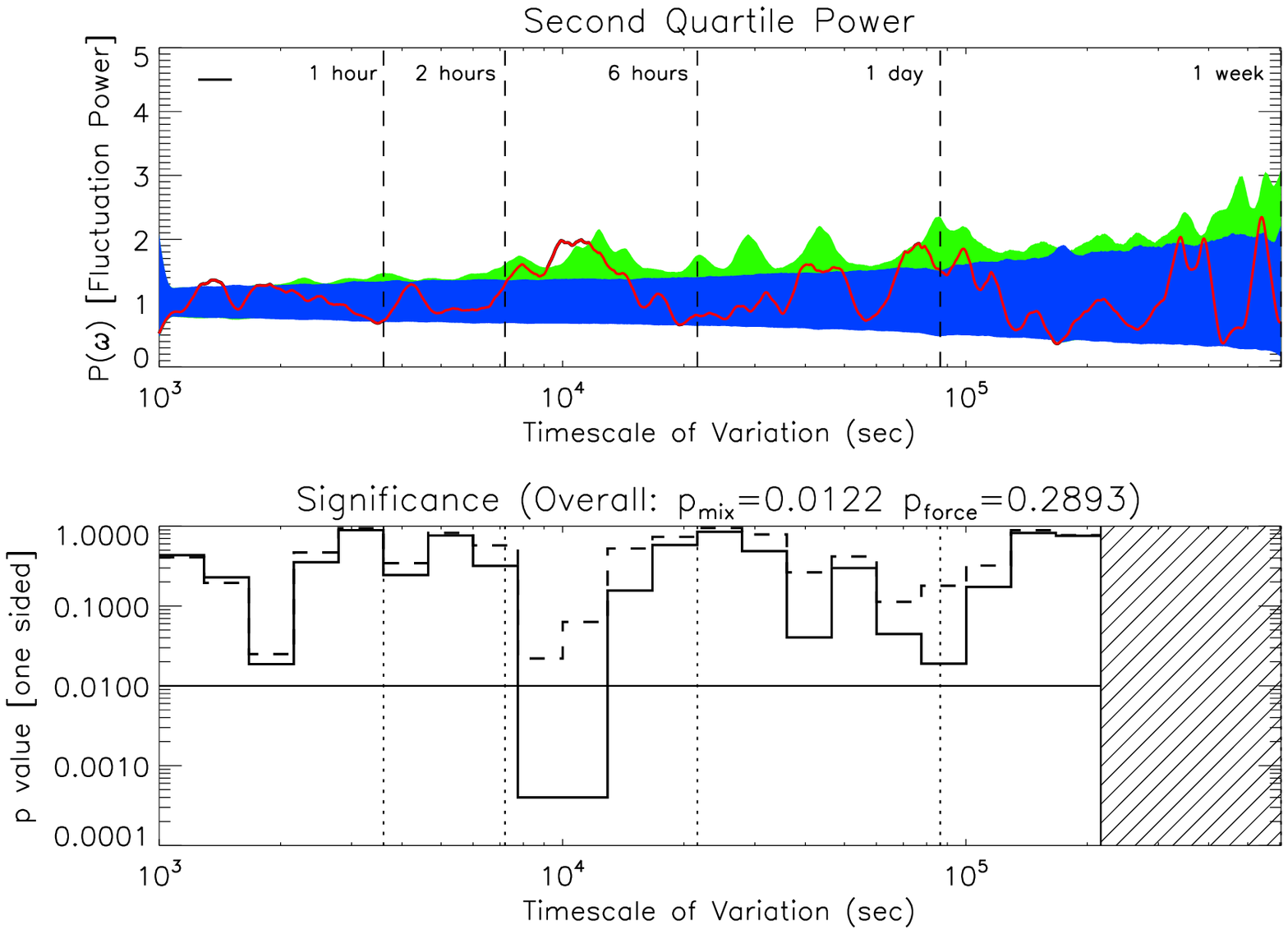} &
\includegraphics[width=3.275in]{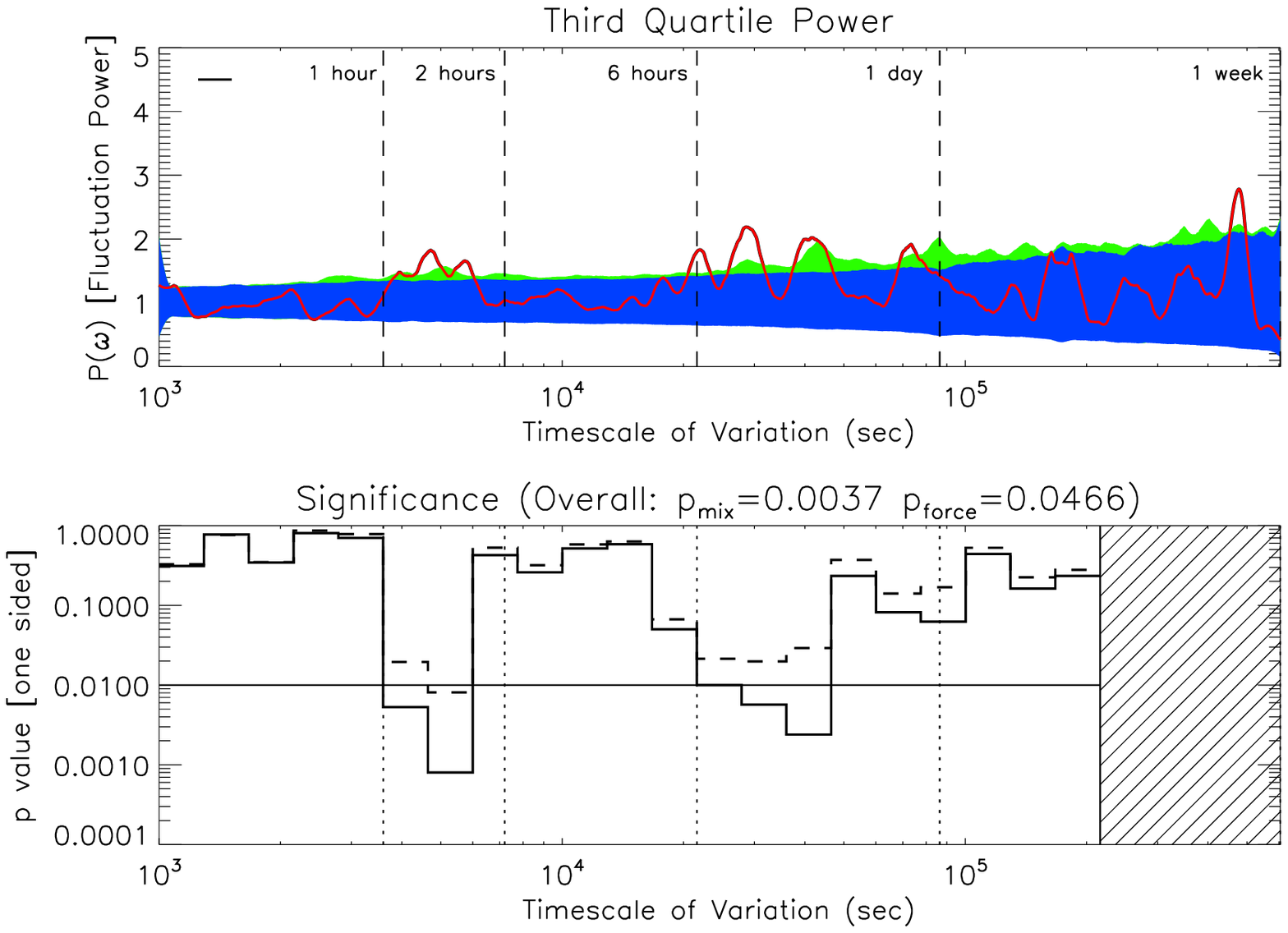}
\end{tabular}
\caption{Uncovering strategic timescales for power and policers. Top left pair: periodogram, and $p$-values, for the policing class -- four top-quartile power individuals who effectively intervene impartially, and break up, conflicts. Top right pair: similarly, for the top quartile in power \emph{minus} policers ($n=8$.) Bottom left pair, similarly for the second power quartile ($n=12$). Bottom right pair: for the third power quartile ($n=12$). In all cases, the demographic classes show significant deviations from the mixed-strategy null. In the policing class, and in the top power quartile, there is also significant deviation from daily forcing, suggesting that the timing of the decision to join or avoid a fight is strategic. There is also evidence of strategic timing, in particular, at the $\alpha$-scale oscillations seen for the 47 socially-mature individuals, when treated collectively  (Fig.~\ref{adults}).} \label{function} \end{figure}

To demonstrate the importance of the daily forcing null, consider the top panel of Fig.~\ref{adults}, which shows the periodogram of the 47 socially-mature individuals considered collectively. There are strong signals of timescales in the $\alpha$ band (around 1 hour) and in the $\gamma$ bands (at 11 hours and 24 hours) -- sufficiently strong that the mixed-strategy null is ruled out at high confidence ($p\aplt0.001$). However, the data for these 47 animals are fully consistent with the daily forcing null, which allows for a changing ``mood'' that systematically shifts the probabilities of becoming involved in fights over the course of the day. Only one individual from the 47 socially mature adults shows significant ($p<0.01$) deviations from the daily forcing null when patterns of behavior are independently examined.

The strongest deviations from daily forcing are found in some of the most important demographic classes in the study group. The top left pair of panels in Fig.~\ref{function} show the timescales associated with the policing class. In addition to the longer $\gamma$ scales, there is strong evidence for timescales, over and above that of daily forcing, that between 1 hour and 2 hours. We return to these strategic time signatures in Sec. \ref{strategic}. 

These results suggest that context-free models are adequate explanations for the collective decision-making pattern exhibited by the 47 socially-mature individuals, but fail as descriptions for the more specific decision-making patterns exhibited by other demographic classes.

Whereas violations of the daily-forcing null indicate strategic timing, it is also worth noting that failure to reject the daily forcing null does not necessarily mean the absence of strategy. Individuals and demographic classes structuring their conflict behavior in response to others whose behavior is driven by daily forcing would show patterns consistent with this rather stringent null and would result in a false negative (Type II error.) Even behavior consistent with the mixed null may have strategic aspects invisible to this analysis, if the relevant contexts are uncorrelated with the size of involved demographic class. \emph{Hence, whereas rejection of the daily forcing null indicates a strategic timescale, failure to reject the daily-forcing null does not necessarily mean the absence of a strategic timescale.} 

\subsection{Timescales for Strategic Behavior}
\label{strategic}

The possibility of strategic decision-making over and above daily forcing is implied in the short timescales of the policing class, which is shown in the top left panels of Fig.~\ref{function}.  These scales are much shorter than can be accounted for by multi-day shifts in behavior. This policing timescale appears in the fastest, $\alpha$ band. As shown in Table~\ref{summary} and Fig.~\ref{sex}, the adult female demographic class also shows strong evidence for violations of the daily-forcing null in the $\alpha$ band.

The timescales of conflict-related activity can thus be mapped onto demographic classes playing functional -- in our case, conflict management -- roles as well as biological and other socially defined demographic classes. The rest of Fig.~\ref{function} shows schematically how individuals with differing social power (in a system in which the power distribution is heavy-tailed \cite{Flack2006b}) show context-sensitive timescales of variation. Lower-power classes show both long and short timescales; the second power quartile shows a $\beta$-scale oscillation at $\sim3$ hours, consistent with daily forcing. In the Materials and Methods, Fig.~\ref{function_supplementary} we show as well the timescales for the fourth and lowest power quartile, which shows marginal evidence for fluctuations at the slower $\beta$-scales. 

\subsection{Summary Table}

\begin{table}
\begin{tabular}{l|c|c|c}
Demographic Class & Timescales & Timescales  & Overall Significance \\ 
& Mixed-Strategy & Daily Forcing & ($10^3$~sec to $2.5$~days) \\
& ($p<0.01$) & ($p<0.01$) & \\ \hline
All Socially-mature Individuals & $\alpha$, $\gamma$ & -- & $p_{\mathrm{mix}}=0.002$; $p_{\mathrm{force}}>0.01$ \\ \hline
\emph{Age  \& Sex} & & & \\
Subadult Females & $\gamma$ & $\gamma$ & $p_{\mathrm{mix}}=0.001$; $p_{\mathrm{force}}>0.01$ \\
Subadult Males & $\alpha$ & - & $p_{\mathrm{mix}}>0.01$; $p_{\mathrm{force}}>0.01$ \\
Adult Females & $\alpha$, $\gamma$ & $\alpha$ & $p_{\mathrm{mix}}=0.01$; $p_{\mathrm{force}}>0.01$ \\
Adult Males & $\alpha$, $\gamma$ & $\gamma$ & $p_{\mathrm{mix}}=0.001$; $p_{\mathrm{force}}=0.01$ \\ \hline
\emph{Social Power \& Role} & & & \\ 
Policers & $\alpha$, $\gamma$ & $\alpha$, $\gamma$ & $p_{\mathrm{mix}}=0.001$; $p_{\mathrm{force}}=0.005$ \\ 
Top Quartile (minus Police) & $\gamma$ & - & $p_{\mathrm{mix}}=0.008$; $p_{\mathrm{force}}>0.01$ \\ 
Second Quartile & $\beta$ & -- &  $p_{\mathrm{mix}}=0.01$; $p_{\mathrm{force}}>0.01$ \\ 
Third Quartile & $\alpha$, $\gamma$ & $\alpha$ &  $p_{\mathrm{mix}}=0.004$; $p_{\mathrm{force}}>0.01$ \\ 
Bottom Quartile & -- & -- &  $p_{\mathrm{mix}}>0.01$; $p_{\mathrm{force}}>0.01$ \\ \hline
Matrilines & $\alpha$, $\beta$ & -- & $p_{\mathrm{mix}} \mathrm{(min)} = 0.003$; $p_{\mathrm{force}}>0.01$
\end{tabular}
\caption{Summary of conflict decision-making timescales detected in the study group. Detections in the various bands ($\alpha$, $\beta$, and $\gamma$) are shown for the two null models. In addition, we show the overall significance of the detections. The primary timescales found are $\alpha$ (between 30 minutes and 2 hours) and $\gamma$ (above 6 hours). One demographic class the second power quartile, shows a significant intermediate, $\beta$, scale between 2 hours and 6 hours. Taken collectively, the 47 socially-mature individuals, as well as a number of smaller demographic classes, show significant evidence overall for non-stationary behavior (above the mixed-strategy null); overall violations of the daily-forcing null, indicating fluctuation-sensitive behavior, are rarer and found in only in the adult males and in the policing class. In the eleven matrilines, there are two detections of $\alpha$ and $\beta$ scales; only one matriline has an overall significance against the mixed-strategy null ($p=0.003$.)}\label{summary}
\end{table}

Table~\ref{summary} summarizes our results for the different demographic classes.

\section{Discussion}
\subsection{Evidence for Multiple Timescales}
Periodic behavior at multiple time scales \cite{Foster} is a fundamental feature of biological systems. Biological periodicities range  from cellular activity measured at the scale to milliseconds, through ultraridian cycles measured at time scales of months or years. Many of these cycles derive from fundamental physiological and biochemical processes, and are observed across distantly related taxa \cite{Cassone}. This study presents, to the best of our knowledge, the first evidence for multiple, periodic timescales associated with social decision-making and behavioral patterns in an animal society, and the first empirical study of social systems to demonstrate the existence of periodicities that are not directly coupled to environmental cycles or known ultraridian mechanisms. Rather we find that for particular sets of individuals playing important conflict management roles, the timescale on which they decide to join or avoid fights is \emph{strategic}. By ``strategic'' we mean that, collectively, these individuals time their decision to join fights in response to correlated fluctuations around the mean pattern of conflict decision-making shown by the rest of the group. 
 
We find three main results. First, whereas some demographic classes have no timescale structure at all, a number of classes show well-separated fluctuations at either short, $\alpha$, or long, $\gamma$, timescales; intermediate $\beta$ scales are seen only rarely in the demographic classes we consider. Secondly, different demographic classes have different timescale signatures, with, for example, subadult females showing strong $\gamma$-band fluctuations whereas the subadult males show fluctuations on in the $\alpha$ band. Finally, and most strikingly, we find a strategic timescale associated with a functional role: a subset of individuals who perform effective policing show a conflict decision-making timescale that is tuned to the mean pattern of conflict in the group and is on the order of one to two hours.

The longest time scales, the $\gamma$ band, are largely but not completely driven by external or internal systematic periodicities, like day-night cycles, feeding cycles, \oe strus, and context-independent fatigue that accrues over the course of a day. These scales are not observed for all group members. The adult females show the clearest day-scale, or $\gamma$, periods in their behavior. 

The $\gamma$ activity in the subadult females that is absent in the subadult males suggests some intriguing sex-related differences in conflict decision-making. The males appear to behave more randomly (in so far as they, as a group, are indistinguishable from the homogeneous mixed-strategy) than the females. Females manifest systematic variation in their willingness to join fights over the course of days. Males on average appear to be more opportunistic in their decisions to joint conflicts in that their decisions are time invariant. These results are consistent with the data on opportunistic coalition formation in males in several primate species \cite{Frans1984,Silk1992}.

The $\beta$ activity on multi-hour scales seen in some of the intermediate- and low-social power groups could relate to simple daily variation in mood, variable sensitivity to hidden triggers in the environment, or group-level variability in temperament that manifests in a variety of behaviors, including conflict-related behavior. 

Of particular interest are the $\alpha$-scale behaviors that can not be explained by daily forcing. On these shorter timescales, we find demographic class periodicities associated with the management of conflict by powerful individuals. These policers appear regularly in the time series at timescale of an hour to two hours.

This result is consistent with previous results showing policers preemptively forestall the escalation of aggression by checking conflicts through impartial interventions \cite{Flack2005b, KrakauerInReview}. It appears that the policers dampen conflict not only by intervening in the regular cycles of fighting, but also by dampening fluctuations about these cycles by making regular appearances in fights. That policing has a signature timescale raises the interesting question of whether policing is predictable by individuals in the society. If so, individuals might be able to tune their conflict decision-making strategies to avoid or facilitate intervention by policers.

\subsection{Implications and Significance}

A body of work in neuroscience \cite{koch1996,Buzsaki:2004p17637}, and preliminary results from the study of social niche construction in animal societies \cite{Boehm2010,Flack2010}, suggest that multiple timescales \emph{within} what is typically considered a ``level'' (e.g. the ``neural level'' or the ``behavioral level'') play an important role in the collective construction of aggregate patterns as well as in inter-individual coordination during communication \cite{Ghazanfar2010}. However, beyond the neural level and the study of biological rhythms, where multiple timescales are well documented, little is quantitatively known about the number, distribution, or significance of timescales. 

This is particularly true at the behavioral level, where there has been little explicit consideration of the role of time in structuring social interactions or in constraining or facilitating the emergence of coordinated aggregates. An important exception is the study of spatial patterning in groups, such as schooling in fish or flocking in birds \cite{couzin2009}. Our analysis differs from these studies in that it stresses periodic variability in a strategic state space rather than non-periodic variability in an explicitly spatial domain. 

There are many important timescale-related questions in the study of social evolution, and many of these concern whether such scales are non-functional emergent properties of collective dynamics, or functional features that serve to better coordinate complex societies. If timescales are functional, how do individuals influence the timescales of behavior of a large group? This study provides provisional evidence that policers, for example, function to modify aggression in the group by performing policing interventions \emph{and} appearing in fights at regular, predictable intervals. This explanation is consistent with the results of an experiment showing that even though the proportion of fights that receive policing interventions is relatively small, aggression increases when the policing function is disabled by ``knockout'' of the policers  \cite{Flack2005}. 

The question of why social and other systems display a range of timescales, as opposed to simpler cases where a single strongly coherent oscillation -- such as the circadian rhythm -- dominates a system, is also of a great interest. It can indicate, among other things, the presence of spin-glass behavior \cite{Coolen:1994p18126}. Near-critical spin-glass properties have been found in the dynamics of neural networks \cite{Tkacik:2006p11059}, and behaviors guided by social interactions may have similar properties. Given the hypothesized role of fluctuation-correlated behavior that violates the daily forcing null, it is also of interest that such glassy systems show non-trivial responses to their own internal fluctuations \cite{dedominicis2006}.

Timescale separations due to the emergence of slow variables at the aggregate level -- \emph{e.g}, the emergence of a slowly changing power structure from a network of status signaling interactions -- are thought to be a means for reducing social uncertainty generated by fluctuations at a lower-level in fight outcomes  \cite{Flack2007, Boehm2010, Flack2010}. In many cases, timescale variability appears to emerge from combinations of connectivity and constraints among populations of components -- in the case of power, for example, this corresponds to a network of individuals signaling about their dominance status. We remain ignorant, however, of mechanisms that might channel variation in timescales at the aggregate level back to influence the timescales on which individuals make decisions. To answer these kinds of questions, we need a means of combining inductive, game-theoretic models of the kind presented in Ref. \cite{DeDeo:2010p18133} with the spectral properties of the highly-resolved behavioral time series as we have presented them here.

\section{Materials and Methods}

Here we cover the methods of data collection protocol, as well as details on the statistical analysis associated with the Lomb-Scargle periodogram and the two null models.

\subsection{Data Collection}

The data collection protocol was approved by the Emory University Institutional Animal Care and Use Committee and all data were collected in accordance with its guidelines for the ethical treatment of nonhuman study subjects.

\subsubsection{Operational Definitions}
\label{op}

Fight: includes any interaction in which one individual threatens or aggresses a second individual. A conflict was considered terminated if no aggression or withdrawal responses (fleeing, crouching, screaming, running away, submission signals) was exhibited by any of the conflict participants for \emph{two minutes} from the last such event. A fight can involve multiple individuals. Third parties can become involved in pair-wise conflict through intervention or redirection, or when a family member of a conflict participant attacks a fourth-party (See Methods). Fights in this data set ranged in size from two to 28 individuals and can be represented as small networks that grow and shrink as pair-wise and triadic interactions become active or terminate, until there are no more individuals fighting under the above described two minute criterion. In addition to aggressors, a conflict can include individuals who show no aggression (\emph{e.g.} recipients or third-parties who either only approach the conflict or show affiliative / submissive behavior upon approaching). Because conflicts involve multiple actors, two or more individuals can participate in the same conflict but not interact directly. 

In this study only information about fight composition (which individuals were involved) and time of fight onset are used. Our analyses focus only on the decision to fight. We do not in this paper consider whether this decision is made with respect to starting a fight or to joining an ongoing fight. We also do not consider any internal aspects of the fight, such as who does what to whom. No time data are available within fights; although the order of an individual's entry was noted, the information was not used in this analysis. The median duration of fights is 15 seconds. The minimum timescale we consider is on the order of 1000 seconds. Given the median duration and this minimum time criterion, deviations between the fight start time and the time of entry of any individual into the fight should not be problematic. 

Fight onset and termination time were noted in hours, minutes, and seconds.  Timing accuracy -- is at worst, on the order of seconds for fight onset time, and so accuracy to this level is more than sufficient for the range of timescales we investigate here.
\\
\\
Demographic Classes: The demographic classes we consider include age-sex classes (see below), matrilines (see below), power quartiles (see below) and policers (see below). 
\\
\\
Age-sex Classes: With the exception of the matriline analyses, all animals in our analyses are ``socially-mature''. Socially-mature males were at least 48 months and socially-mature females were at least 36 months by study start. Subadult males were males between 48 and 60 months; adult males were at least greater than or equal to 60 months. Subadult females were at least 36 months but less than 48 months; adult females were at least greater than or equal to 48 months. These thresholds correspond to approximate onset of social maturity in pigtailed macaques.
\\
\\
Matriline: an adult female and her daughters. In the study group, all females in a matriline were related through the maternal line. Only females one year or older were included in the matriline analyses.
\\
\\
Power: the degree of consensus among individuals in the group about whether an individual is capable of using force successfully. Consensus is quantified by taking into account the total number of subordination signals an individual receives and multiplying this quantity by a measure of the diversity of signals received from its population of signalers (quantified using Shannon Information) \cite{Flack2006b}. In the pigtailed macaque, the subordination signal is the silent bared teeth display \cite{Flack2007}. The distribution of power in our study group is heavy tailed. The first power quartile corresponds to the top 12 individuals of the 48.
\\
\\
Policers: four individuals (one female, three males) who preform the majority of effective policing interventions (sit towards the tail in a log normal distribution of the frequency of effective policing interventions). A policing intervention is an impartial intervention performed by a third party into an ongoing conflict. \cite{Flack2005}. These individuals occupy the top four spots in the power structure and sit toward the tail of the distribution. 
\\
\\
General note about demographic classes: Our results showing that demographic classes that have signature timescales suggests that there are empirical grounds for treating them as coherent units with sets of actions. This is similar to the concept of coalitions in cooperative form games \cite{myerson1991}, and the finding is consistent with results of a previous study in which we showed that the triad, not the individual or the dyad, is a the fundamental unit of conflict dynamics in this group \cite{DeDeo:2010p18133}.

\subsubsection{Data Collection Protocol}
During observations all individuals were confined to the outdoor portion of the compound and were visible to the observer, JF. The $\approx150$ hours of observations occurred for up to eight hours daily between 1,100 and 2,000 hours over a twenty-week period, comprising roughly 122 days, from June through October 1998 and were evenly distributed over the day. This span allows us to study a wide range of scales on which behavior can change. The sampling is sparse relative to the total number of hours (150 of 2928) in the data collection period; it is also irregular, in that observational periods are not separated by the same number of days and have different lengths and gaps. Fight and status signaling data were collected using all-occurrence sampling. 

Provisioning occurred before observations, and once during observations at the same time each day. The group was stable during the data collection period (defined as no reversals in status signaling interactions resulting in a change to an individual's power score, see \cite{Flack2006b}). One animal, Ud, was removed from the group for health reasons towards the end of the study; as this sudden removal (and thus zeroing out of all behavior data) is likely to produce strong, but spurious signals of behavioral variation, we excluded her from the analysis.

\subsection{$Q$ Values and Timescale Coherence}

In common with many spectral analysis methods, Lomb-Scargle takes as basis functions the sine and cosine, phase shifting them to find the optimal fit. A pure sine-wave signal, for example, would amount to an extremely sharp spike at the relevant frequency in the periodogram.

Although precise oscillations are unlikely to be found in the noisy and non-ergodic environments we consider here, more realistic behaviors are also mapped to the relevant portions of the plot; for example, repetitive excitation and subsequent exponential decay would map to a peak centered around the excitation period. If one also allowed there to be jitter in the exponential decay -- random variation both in the time-constant and in the precise timing of the excitations following decay -- the peak would broaden further.

One sometimes defines a $Q$-factor, a measure of the width of a peak in a periodogram, as $f_{0}/\Delta f$; here $f_{0}$ is the peak center and $\Delta f$ the width of the peak at half-maximum. Very ``pure'' oscillations -- close to a sinusoidal variation -- have high $Q$-factors; conversely, purely damped systems that dissipate oscillations -- such as a suspension system in a car or building -- have $Q$ less than unity.

Man-made systems such as optical cavities for lasers can have $Q$ factors in the millions and billions; mechanical vibrating systems such as a tuning fork have $Q$ of of order $10^{3}$. Meanwhile, natural phenomena tend to have much lower $Q$-factors, indicating the presence of noise and blended signals at different scales. For example the $Q$-factors of brain oscillations measured from an EEG of a sleeping human of can be of order 10 or 100;  the quasi-periodic phenomena found in neutron star systems can have $Q$ of order 10 or lower \cite{Middleditch:1986p18179,vanderKlis:1985p18178}.

In the system we study here, we find $Q$ factors of larger than, but of order, unity -- \emph{i.e.}, slightly less coherent than the quasi-periodic oscillations of the human brain. These are similar to the $Q$-factors one can estimate for the bacterial motors of \emph{E. coli} measured in Ref. \cite{Korobkova:2004p17638}, and higher than the $Q$-factors seen for the signaling networks that control them.

\subsection{Statistical Features of Broadband Lomb-Scargle}

The Lomb-Scargle periodogram, introduced in Sec.~\ref{methods}, forms the center of the data analysis in this paper. Traditionally it has been applied to the detection of high-$Q$ signals such as the detection of orbital periods of stellar systems. We discuss here some of the statistical tools employed to uncover much the broader features of the spectrum observed in our study system.

It turns out that despite the discreteness of the measurement values -- and the correlations between conflict behaviors of different individuals at the same time -- the distribution of $P(\omega)$ under the mixed-strategy null of Sec.~\ref{mixed} is also approximately both exponential and of mean unity. The daily forcing null is far more structured, and induces strong correlations between bins that can be seen visually in the plots.

If one is then seeking a signal at a precise frequency -- \emph{i.e.}, a sinusoidal oscillation with an extremely high $Q$-value -- then it becomes simple to determine a threshold power above which a detection is considered significant with a certain $p$-value. The approximate value is \begin{equation} P(\omega)\approx-\ln{p/M},\end{equation}  where $M$ is the number of independent frequencies. The dependence on $M$ comes from the fact that rare events become more likely the more one samples -- in the words of Ref. \cite{press1996}, ``look long enough, find anything!''

However, for signatures that are more broad-band -- that are expected to cover a range of frequencies -- this simple method is too conservative. The presence of correlated noise makes the analytic estimation even harder, since, in contrast to the Gaussian, there is no simple version of the multivariate exponential distribution that allows for correlations between arbitrary numbers of different frequencies. Instead of analytic approximations, then, a Monte Carlo estimate of bin-by-bin significance is made: many instantiations of the null model are produced, and their distribution compared to measured value, bin-by-bin, to produce an estimate of the $p$-value.

Given a set of such $p$-values for all bins, we then wish to estimate, in Sec.~\ref{mixed}, the overall $p$-value for a detection of non-null timescales. The Bonferroni correction for combining $p$-values is generally considered to be too conservative. In a search for periodicity in gene expression data, Ref. \cite{Glynn:2006p17794} used an order statistic on $p$-values. Here, following Ref. \cite{Loughin:2004p18040}, we use the $\chi^{2}(2)$ test of Fisher, which works well where one is seeking evidence for strong signals in one or two bins. All these methods require that the $p$-values be independent (in the null).

Failures of null independence -- and, in general, failures of null $p$-values to fall in a $U(0,1)$ distribution -- can lead to both Type I and Type II errors for any method of combination. We check the validity of the $\chi^{2}(2)$ test by running our same analyses on null data alone. We find that the Fisher method works reasonably well, though not perfectly, and can underestimate $p$-values when $p\leq10^{-3}$. We silently insert these corrections to reported $p$-values in the main paper, so that our stated $p$-values for overall significance are our best estimates, and do not rely on the strong assumptions of combination methods such as the $\chi^{2}(2)$ test.

Our bins are logarithmically spaced; we search the range between $10^{3}$~seconds and $2.5$~days, and, once we correct for correlations, our results are, as they should be, insensitive to the number of bins. The periodograms themselves are very noisy, with fluctuations from point to point; when we show them, we smooth with a small window; the width of this window is shown visually as the short black line in the top left corner of each plot. The smoothing is done by a Savitzky-Golay filter, which helps preserve the sharpness (or lack thereof) of the spectral features and is less likely to bias the amplitude \cite{press1996}. All of our significance estimates are made, naturally, on the unsmoothed data.

In the case of the mixed-strategy null, we found our results insensitive to whether we sampled with replacement or simply shuffled the timeseries (sampling without replacement.) This is a consequence both of how much data we have, and that the Lomb-Scargle technique is not strongly sensitive to very rare events.

In the case of the daily-forcing null, we bin fight sizes for each demographic class in 15 minute increments over the day to produce a distribution that we then draw from to simulate a new series of conflicts. (Results are largely insensitive to bin size; in many cases, the daily-forcing null can be replaced with a parametrized fit -- for example, the overall conflict behavior of the study group can be modeled as a Poisson process with a mean that slowly increases over the course of the day.)

\subsection{Periodograms}

\begin{figure} 
\begin{tabular}{cc}
\includegraphics[width=3.275in]{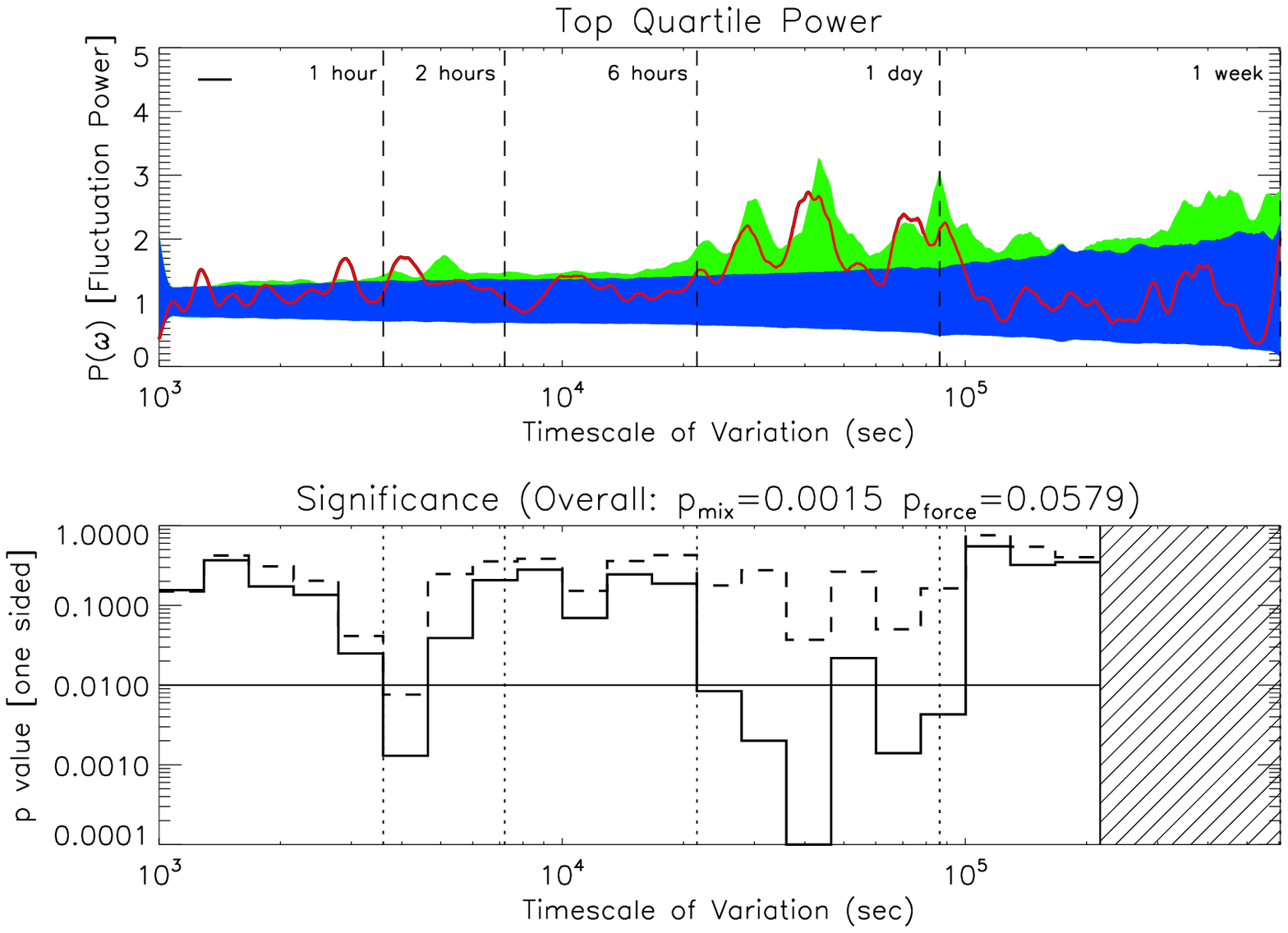} \\
\includegraphics[width=3.275in]{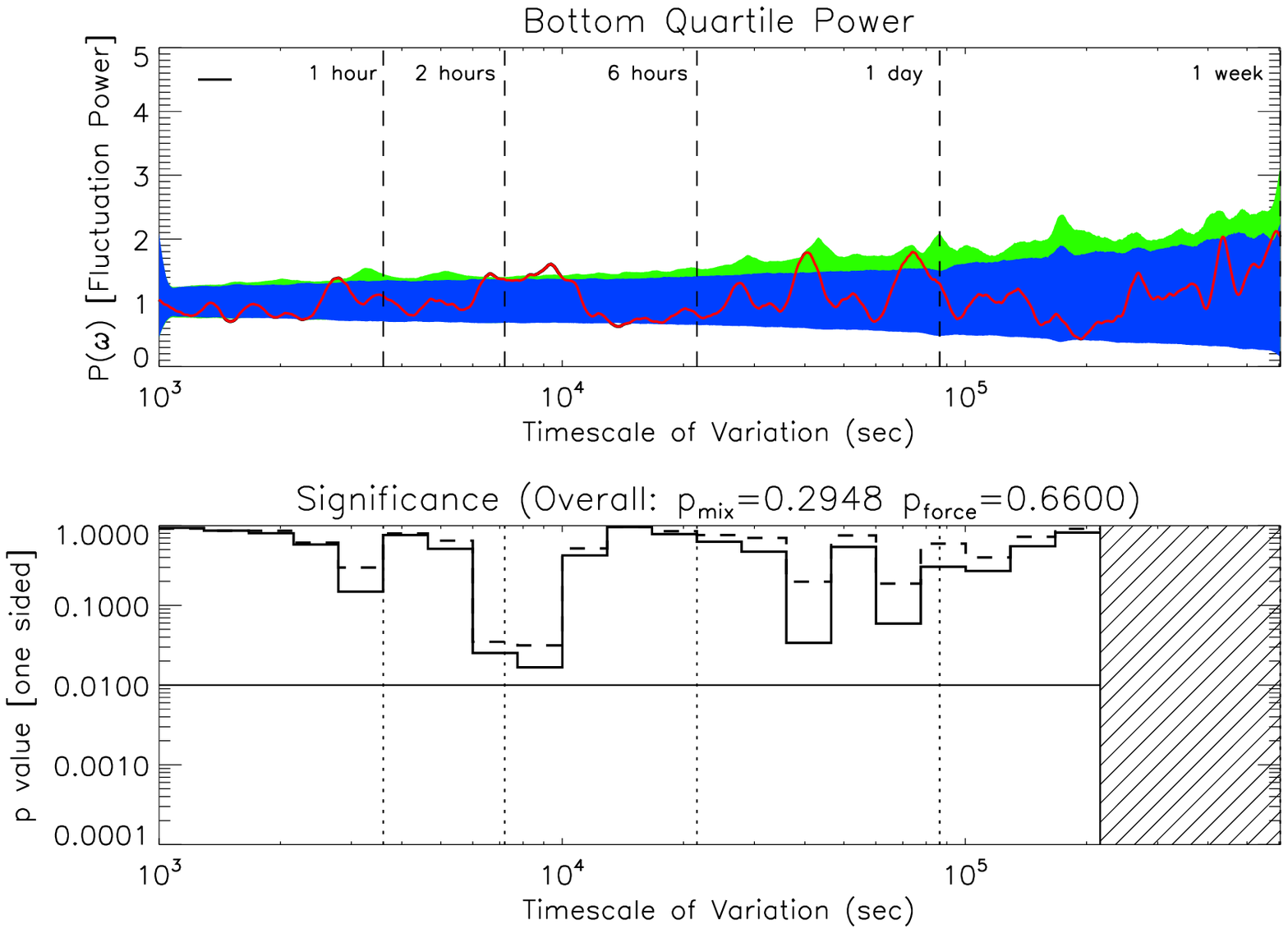}
\end{tabular}
\caption{Top pair: the top quartile in power, \emph{including} police class ($n=12$; in the main text this quartile is split to show the influence of the policing functional class.) Bottom pair: the lowest quartile in power ($n=11$.)} \label{function_supplementary}
\end{figure}

\begin{figure} 
\begin{tabular}{cc}
\includegraphics[width=3.275in]{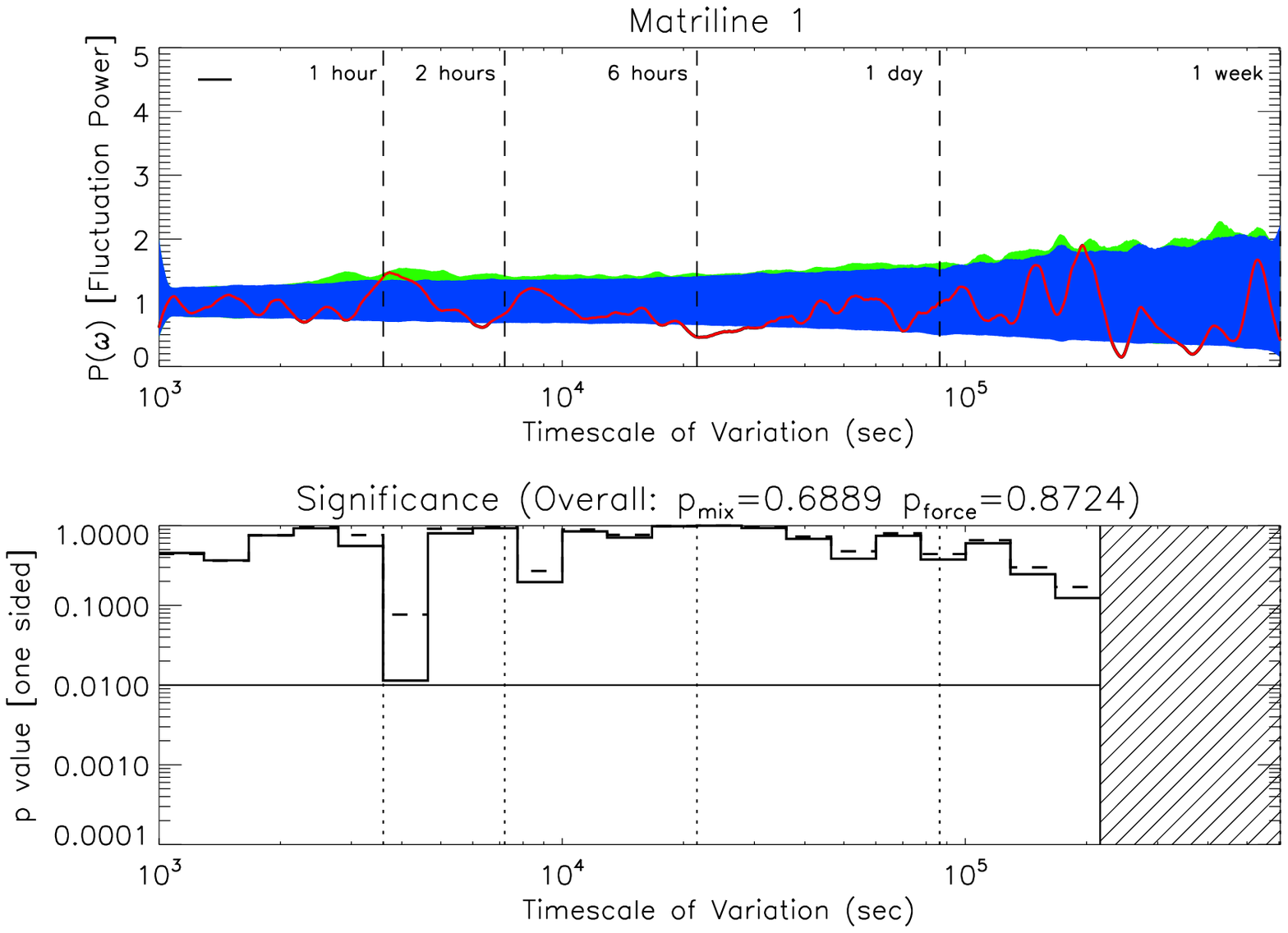} &
\includegraphics[width=3.275in]{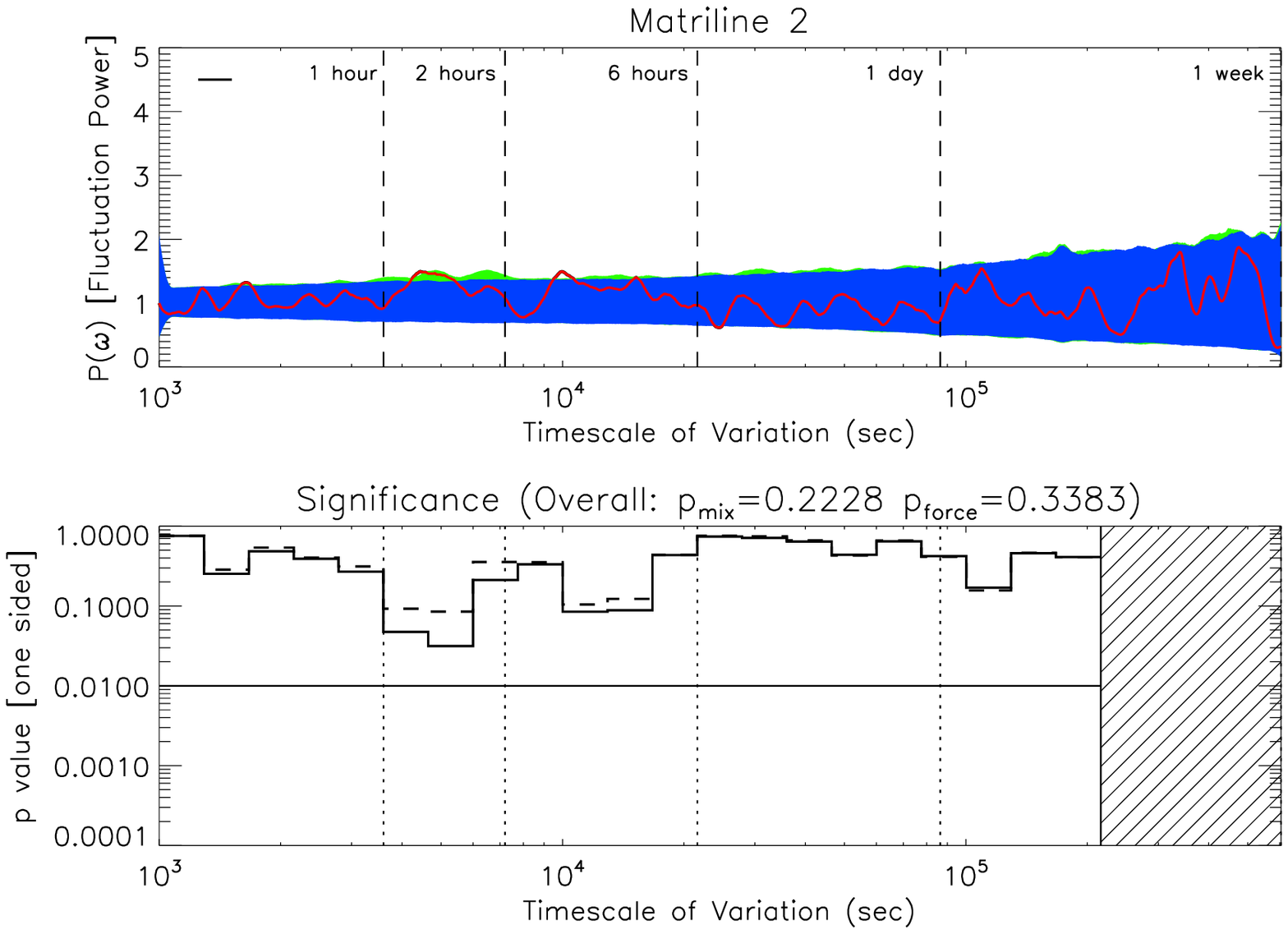} \\
\includegraphics[width=3.275in]{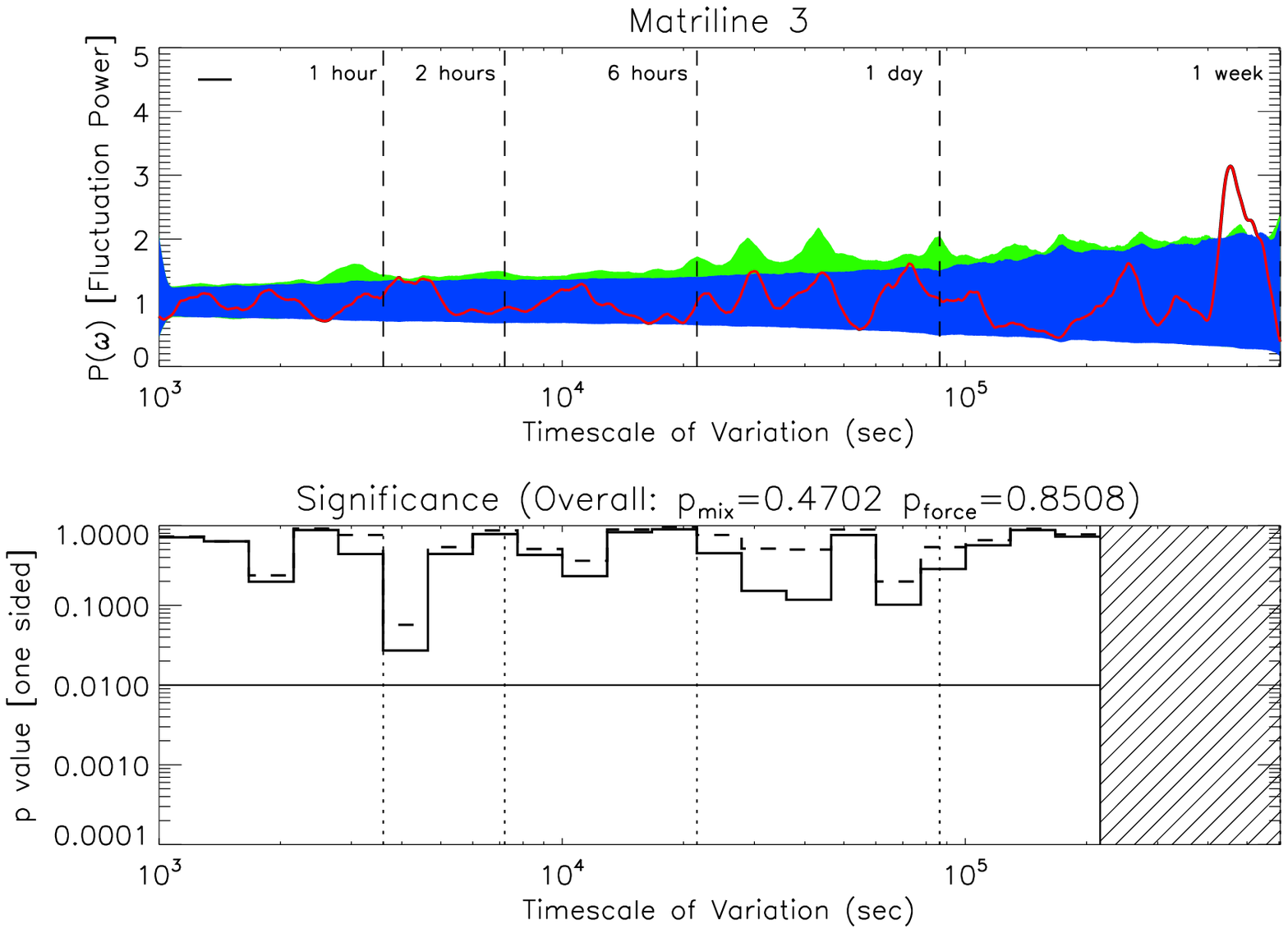} &
\includegraphics[width=3.275in]{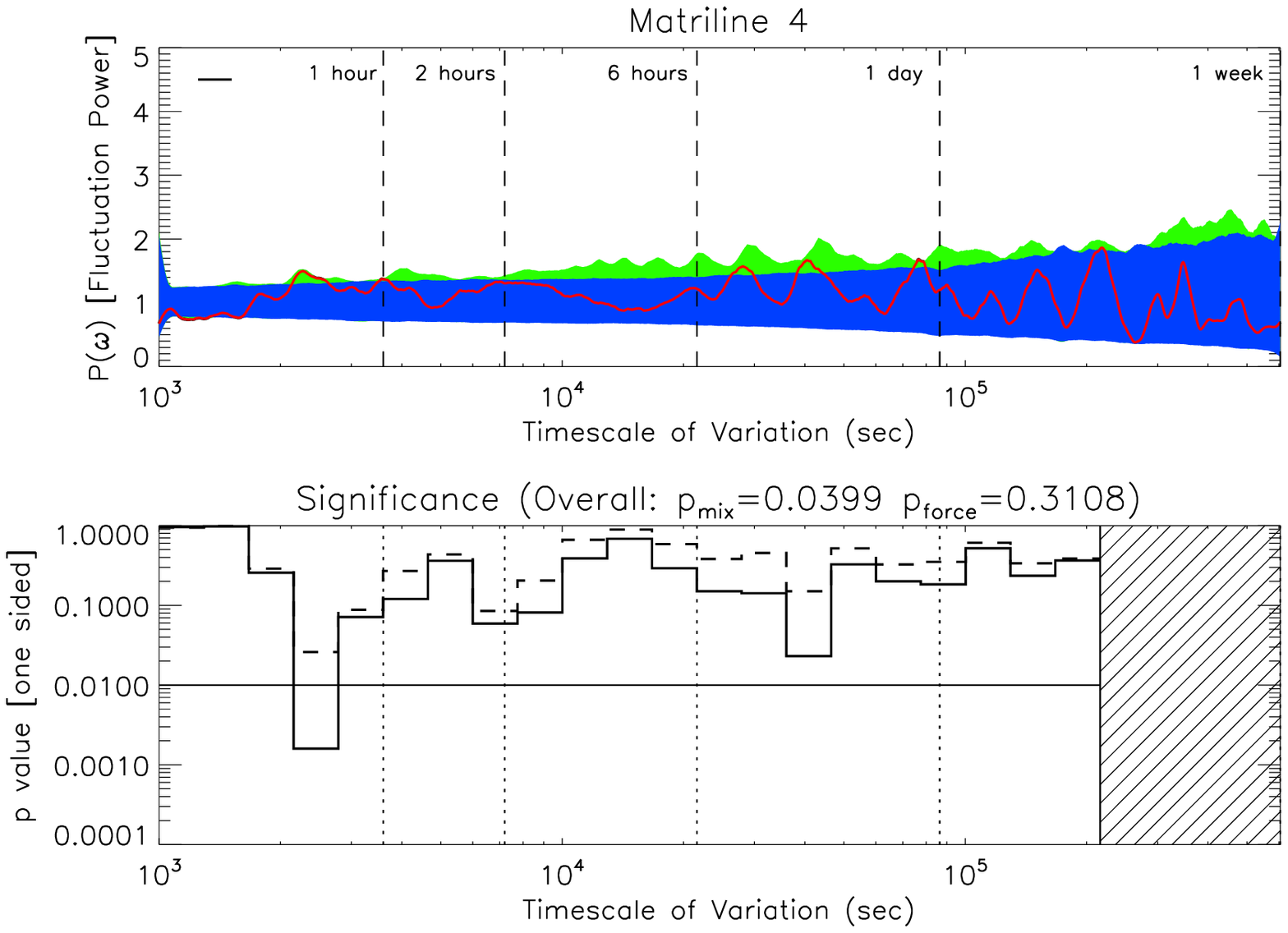} \\
\end{tabular}
\caption{Matrilines 1 through 4. $n=\{4,3,3,2\}$} \label{m1_supplementary}
\end{figure}

\begin{figure}
\begin{tabular}{cc}
\includegraphics[width=3.275in]{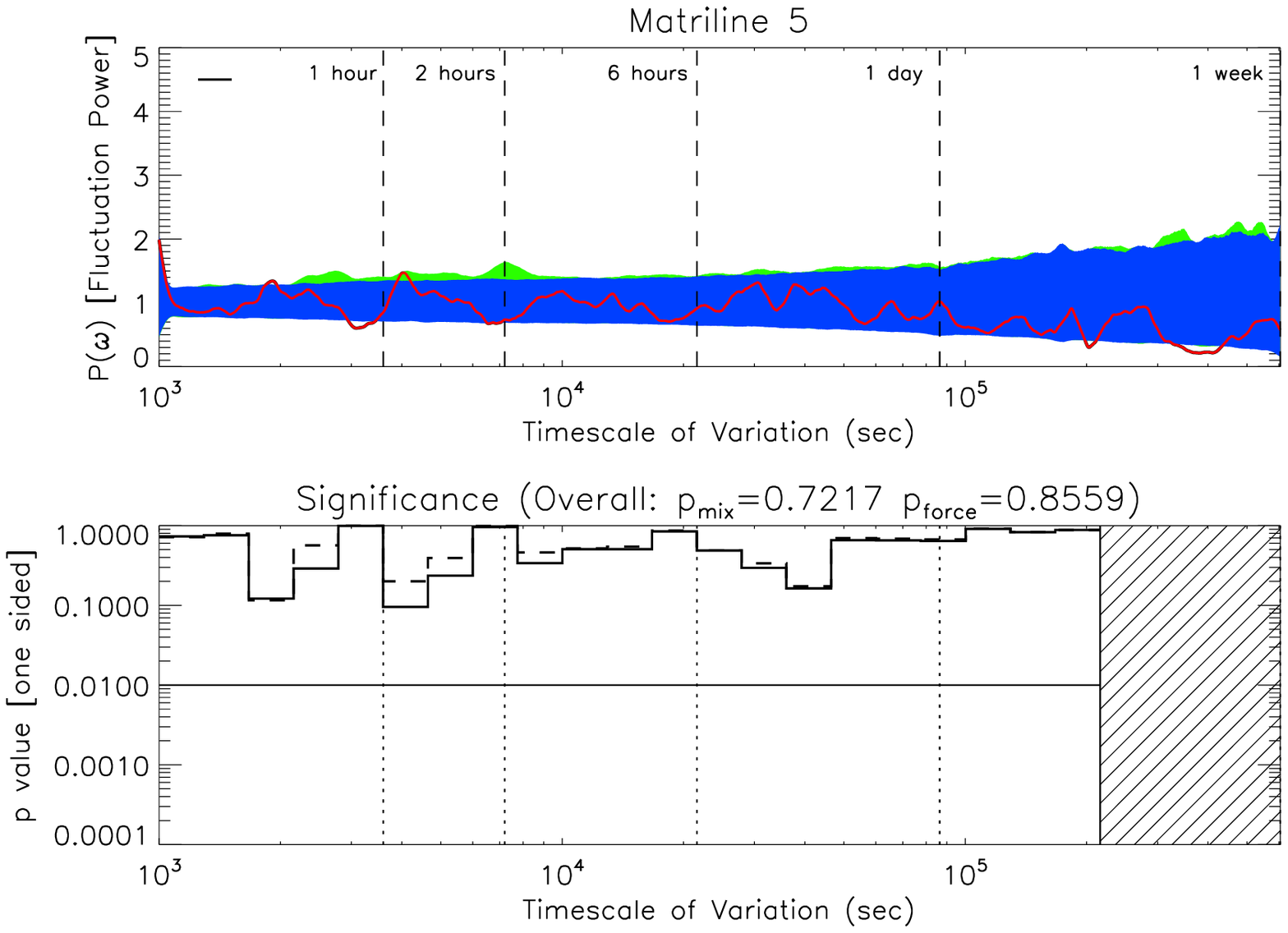} &
\includegraphics[width=3.275in]{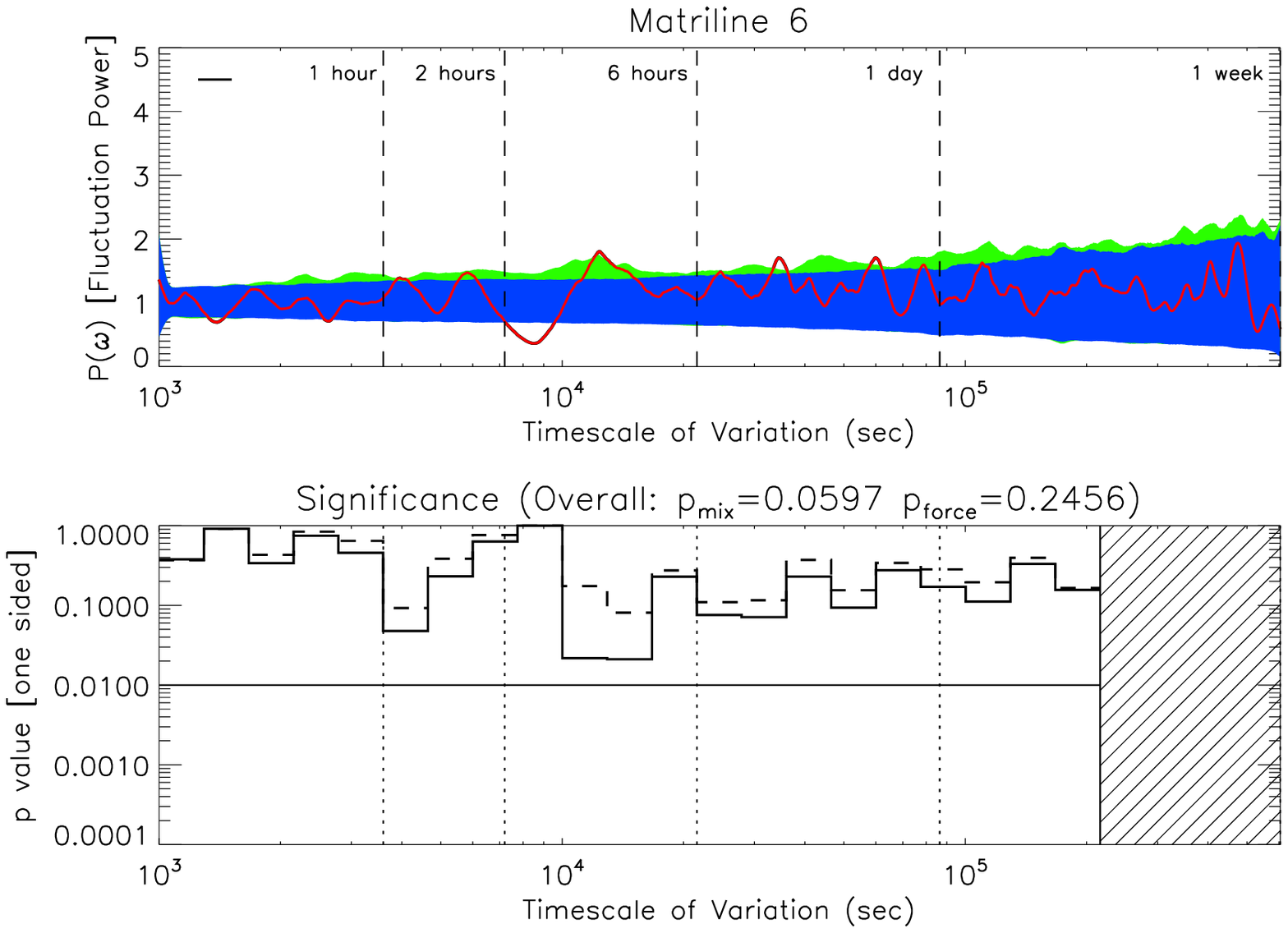} \\
\includegraphics[width=3.275in]{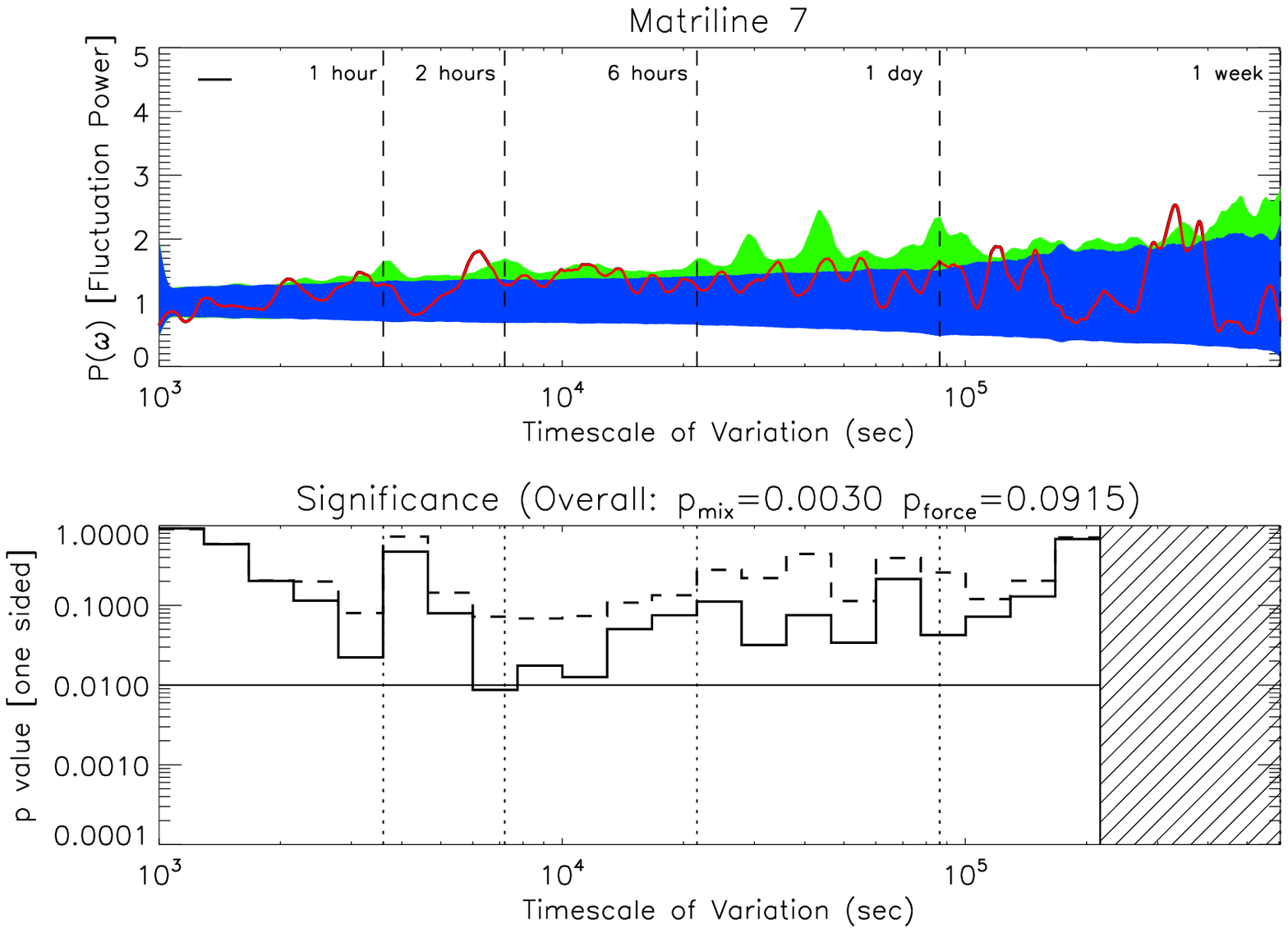} &
\includegraphics[width=3.275in]{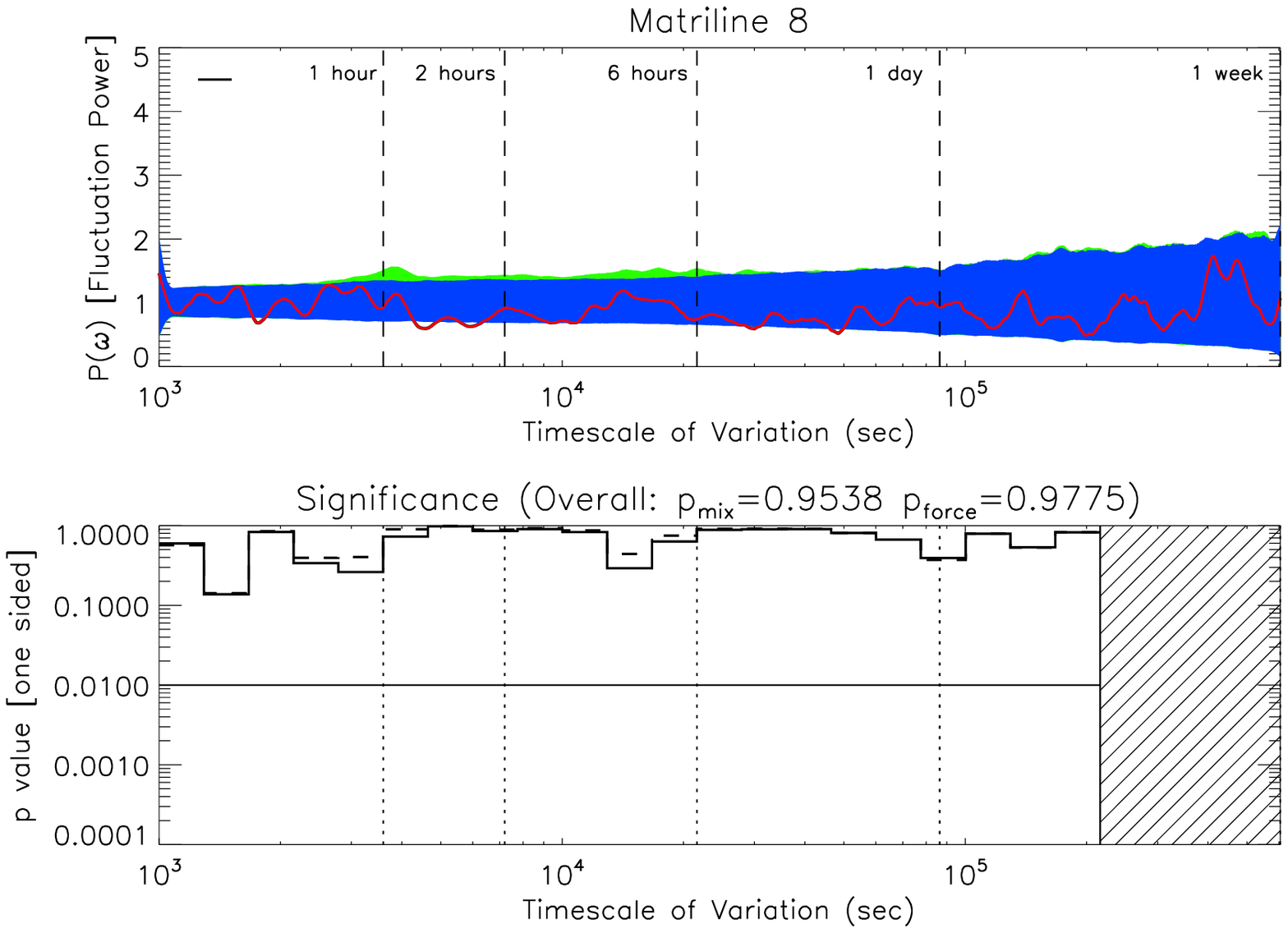} \\
\end{tabular}
\caption{Matrilines 5 through 8. $n=\{3,2,3,3\}$} \label{m2_supplementary}
\end{figure}

\begin{figure}
\begin{tabular}{cc}
\includegraphics[width=3.275in]{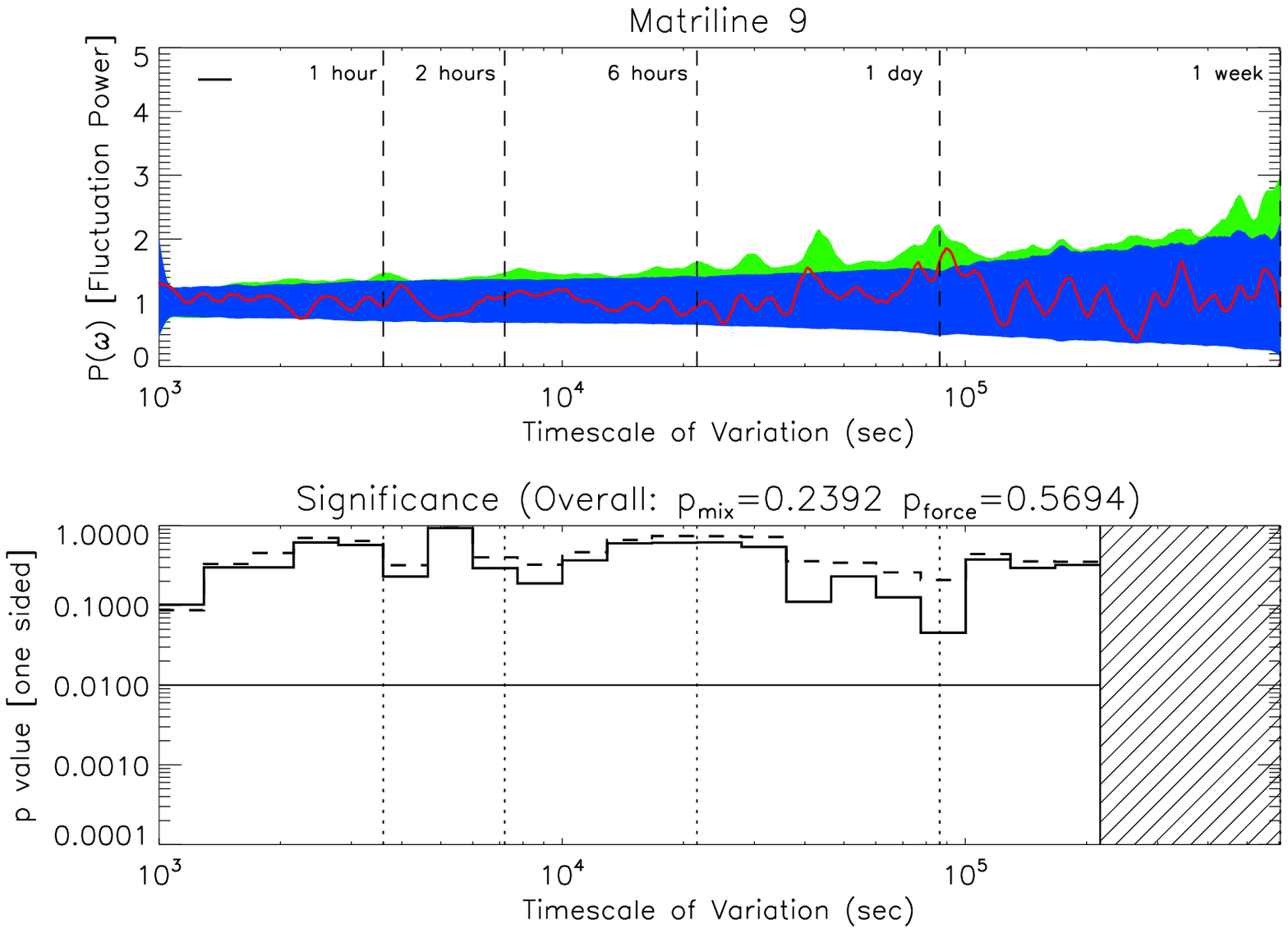} &
\includegraphics[width=3.275in]{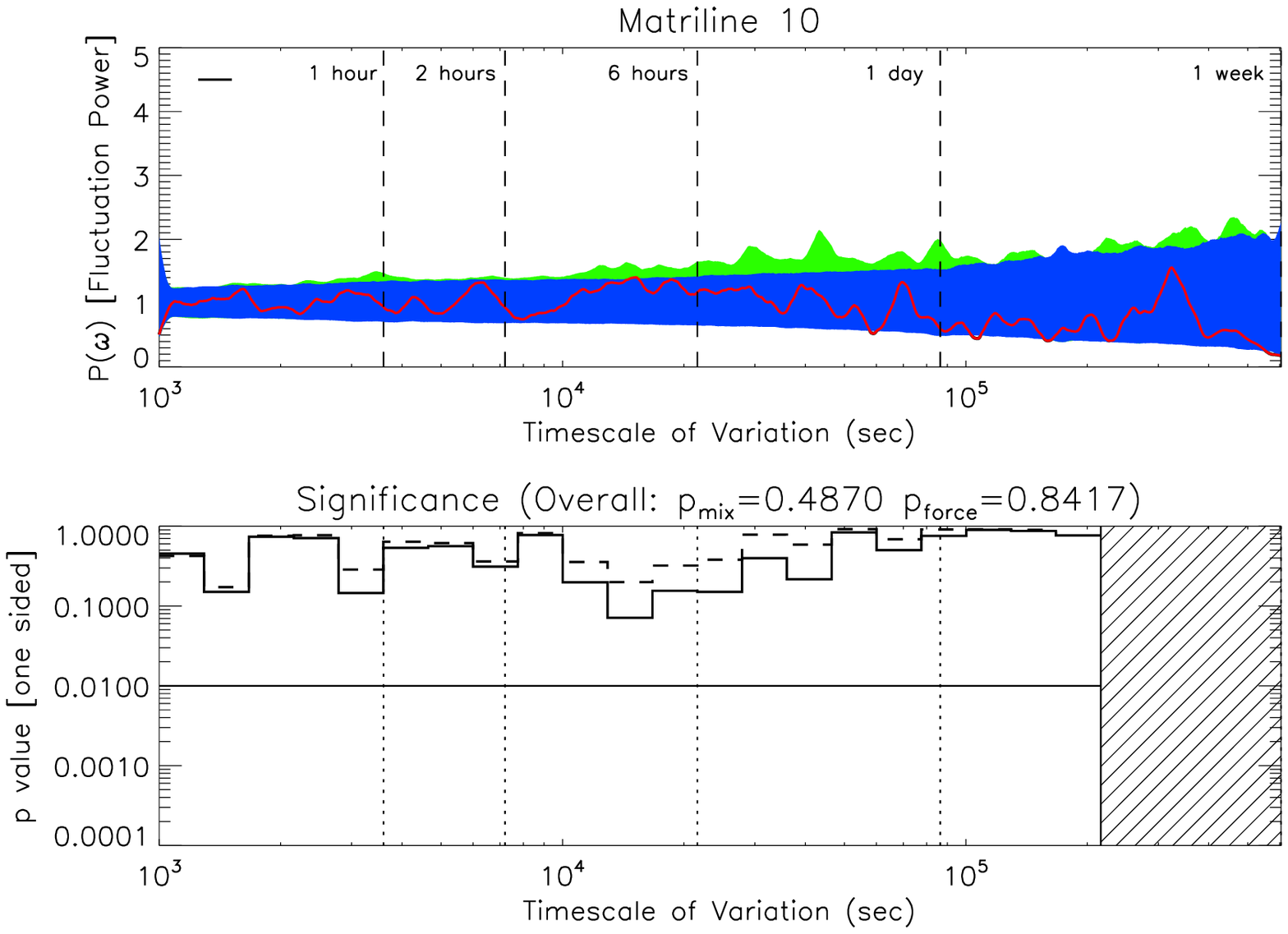} \\
\includegraphics[width=3.275in]{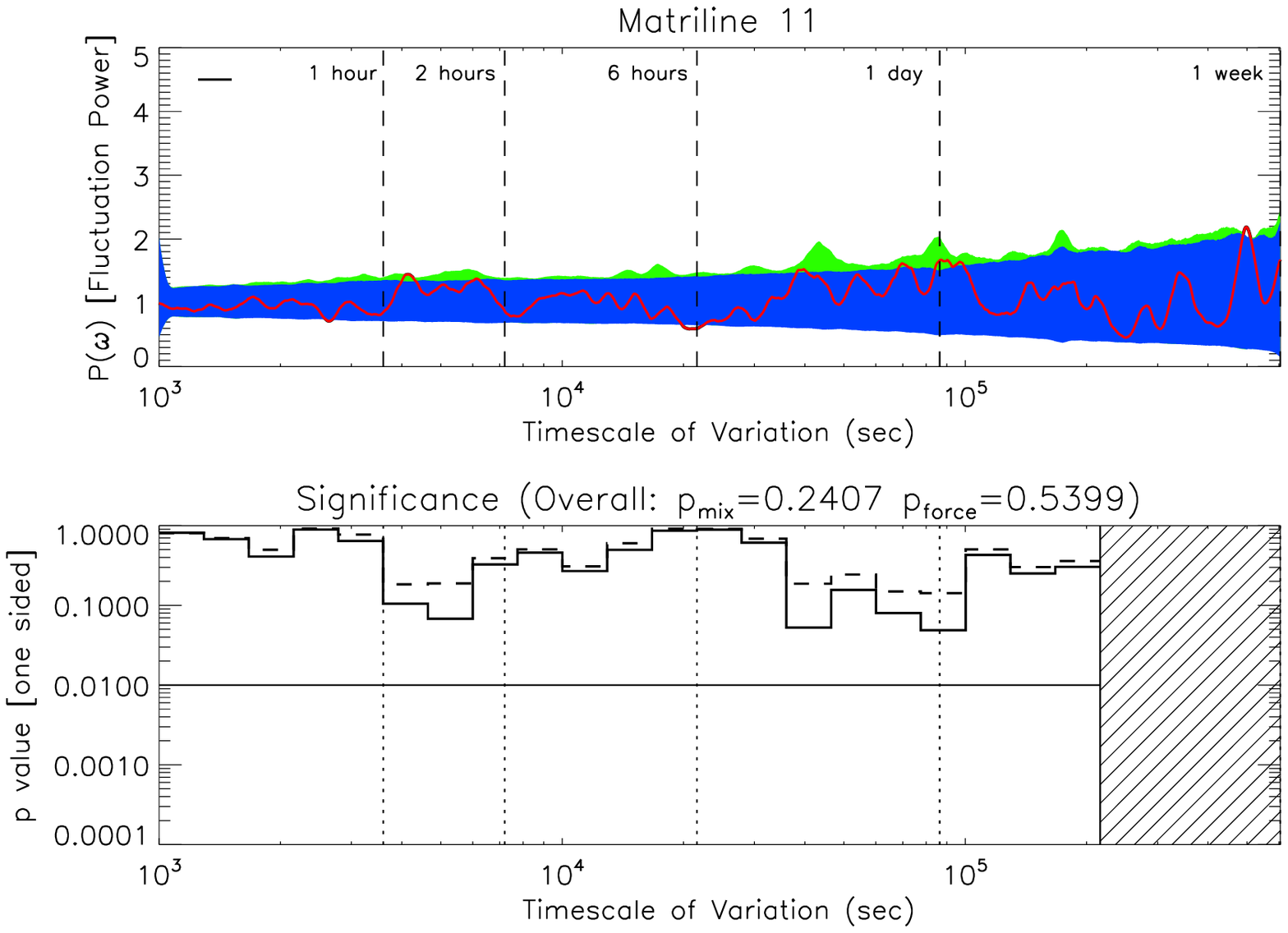} \\
\end{tabular}
\caption{Matrilines 9 through 11. $n=\{3,2,2\}$} \label{m3_supplementary}
\end{figure}

In Figs.~\ref{function_supplementary} through~\ref{m3_supplementary}, we show the periodograms for all the remaining demographic classes discussed in the paper. These results are also summarized in Table~\ref{summary}.

\clearpage

\section{Acknowledgements}

We thank Frans de Waal for support during data collection, and the staff of the Yerkes National Primate Research Center, for help with the data collection.

\end{document}